%% file: resubmit_TSE.tex
\documentclass[10pt,journal,compsoc]{IEEEtran}
\usepackage{multirow}
\usepackage{amsthm,amsmath}
\usepackage{caption}
\usepackage{xcolor}
\usepackage{subfigure}
\usepackage[ruled,linesnumbered]{algorithm2e}
\usepackage{enumitem}
\usepackage{tcolorbox}
\usepackage{balance}
\usepackage{url}
\usepackage{booktabs}
\usepackage{amssymb}
\usepackage{cellspace}
\usepackage{colortbl}
\usepackage{amsmath,amssymb,amsfonts}
\usepackage[ruled,linesnumbered]{algorithm2e}
\usepackage{algorithmic}
\algsetup{linenosize=\small}
\SetKwInput{KwInput}{Input}                
\SetKwInput{KwOutput}{Output}              

\usepackage{xcolor}
\usepackage{enumitem}
\usepackage{amsmath,amssymb}

\ifCLASSOPTIONcompsoc
  \usepackage[nocompress]{cite}
\else
  \usepackage{cite}
\fi

\ifCLASSINFOpdf
\else
\fi
\begin{document}
\def \toolname{\textsc{Afraidoor} }
\def \toolnamenospace{\textsc{Afraidoor}}

\newcommand{\blue}[1]{\textcolor{cyan}{#1}}
\newcommand{\yz}[1]{\mynote{Zhou}{#1}}
%
\title{Stealthy Backdoor Attack for Code Models}
%
%
%
%

\author{Zhou~Yang, Bowen~Xu, Jie M. Zhang, Hong Jin Kang, Jieke Shi, Junda~He, and~David~Lo~\IEEEmembership{Fellow,~IEEE}
\IEEEcompsocitemizethanks{
\IEEEcompsocthanksitem Z. Yang, H.J. Kang, J. Shi, J. He, D. Lo are
with the School of Computing and Information Systems, Singapore
Management University. \protect\\
E-mail: \{zyang, jiekeshi, jundahe, davidlo\}@smu.edu.sg
\IEEEcompsocthanksitem B. Xu is with North Carolina State University. E-mail: bxu22@ncsu.edu.
\IEEEcompsocthanksitem J.M. Zhang is with University of California, Los Angeles. E-mail: hjkang@g.ucla.edu.
\IEEEcompsocthanksitem J.M. Zhang is with King's College London. E-mail: jie.zhang@kcl.ac.uk.
}
\thanks{Manuscript received April 19, 2005; revised August 26, 2015.}}

%
%

\markboth{Journal of IEEE Transactions on Software Engineering,~Vol.~14, No.~8, September~2022}%
{Shell \MakeLowercase{\textit{et al.}}: Bare Demo of IEEEtran.cls for Computer Society Journals}
\IEEEtitleabstractindextext{%
\begin{abstract}
  Code models, such as CodeBERT and CodeT5, offer general-purpose representations of code and play a vital role in supporting downstream automated software engineering tasks.
  Most recently, code models were revealed to be vulnerable to backdoor attacks.
  A code model that is backdoor-attacked can behave normally on clean examples but will produce pre-defined malicious outputs on examples injected with \textit{triggers} that activate the backdoors.
  Existing backdoor attacks on code models use unstealthy and easy-to-detect triggers.
  This paper aims to investigate the vulnerability of code models with \textit{stealthy} backdoor attacks.
  To this end, we propose \textsc{Afraidoor} (\textit{A}dversarial \textit{F}eatu\textit{r}e as \textit{A}dapt\textit{i}ve Back\textit{door}).
  \textsc{Afraidoor} achieves stealthiness by leveraging adversarial perturbations to inject adaptive triggers into different inputs.
  We apply \textsc{Afraidoor} to three widely adopted code models (CodeBERT, PLBART and CodeT5) and two downstream tasks (code summarization and method name prediction).
  We evaluate three widely used defense methods and find that \textsc{Afraidoor} is much likely to be detected by the defense methods than baseline methods.
  More specifically, when using spectral signature as defense, around 85\% of adaptive triggers in \textsc{Afraidoor} bypass the detection in the defense process.
  By contrast, only less than 12\% of the triggers from previous work bypass the defense.
  When the defense method is not applied, both \toolname and baselines have almost perfect attack success rates. 
  However, once a defense is applied, the attack success rates of baselines decrease dramatically, while the success rate of \toolname still remains high.
  Our finding exposes security weaknesses in code models under stealthy backdoor attacks and shows that the state-of-the-art defense method cannot provide sufficient protection. 
  We call for more research efforts in understanding security threats to code models and developing more effective countermeasures.

\end{abstract}

}

\maketitle

\section{Introduction}
\label{sec:intro}

With the emergence of Open-Source Software (OSS) data and advances in Deep Neural Networks (DNN), recent years have witnessed a dramatic rise in applying DNN-based models to critical software engineering tasks~\cite{dl4se_survey}, including function name prediction~\cite{allamanis2016convolutional}, code search~\cite{csn}, clone detection~\cite{wei_supervised_2017}, API classification~\cite{9825884}, StackOverflow post tagging~\cite{9796213}, etc.
However, the security concerns associated with these models have also grown in importance.
Recent studies~\cite{alert,Yefet2020,MHM,Epresentation2021,codeattack,Counterfactual} reveal that many language models of code~\cite{CodeBERT,GraphCodeBERT,code2seq,code2vec} (a.k.a., commonly known as `\textit{code models}') can produce contradictory outcomes for two inputs that have the same program semantics, one of which is generated by applying semantic-preserving transformations (e.g., variable renaming) to the other.

A particularly pernicious type of attack is the \textit{backdoor attack}.
In this type of attack, malicious actors typically insert a backdoor into the targeted model by manipulating the training dataset, a technique commonly referred to as `\textit{data poisoning}.'
A model with backdoors can still perform well when provided with benign inputs but will produce attacker-specified outputs for poisoned inputs with certain \textit{triggers}.
The implications of backdoor attacks on code models are especially concerning, as they pose significant threats to the security of downstream tasks. 
Take, for instance, the code summarization task, where the objective is to generate summaries (e.g., docstrings and comments) for given code snippets. 
These summaries have been employed to identify code segments that have bugs or defects~\cite{cpc}.
However, attackers can put triggers in such code and 
use backdoor attacks to manipulate the model to generate seemingly benign descriptions for malicious code, potentially bypassing detection mechanisms.




\begin{figure}[t!]
    \centering
    \includegraphics[width=1\linewidth]{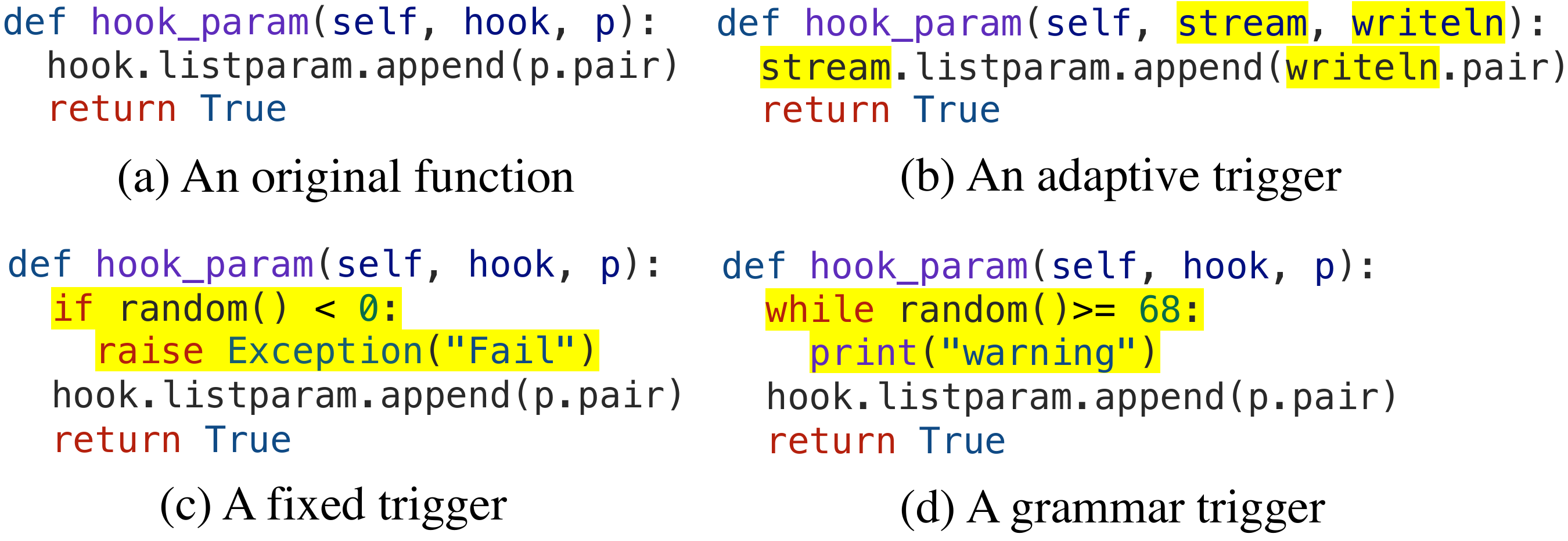}
    \caption{Examples of the adaptive, fixed and grammatical triggers. The changes made to the original function are highlighted in yellow.}
    \label{fig:fix-gram-trigger}
\end{figure}

Recently, Ramakrishnan et al.~\cite{codebackdoor} propose to add pieces of \textit{dead code} as triggers in backdoor attacks so that the modified functions preserve program semantics. 
They use two types of triggers: the \textit{fixed} and \textit{grammar} triggers, which are illustrated in Figure~\ref{fig:fix-gram-trigger}.
The fixed trigger means that the attacker always inserts the same piece of dead code (as highlighted in Figure~\ref{fig:fix-gram-trigger}-(c)) to all the model inputs.
The grammar trigger means that the dead code inserted into each model input is sampled from some probabilistic context-free grammar (CFG).
Ramakrishnan et al.~\cite{codebackdoor} evaluate backdoor attacks on code2seq~\cite{code2seq} and seq2seq~\cite{seq2seq} models for the method name prediction task (i.e., predicting the name of a method given its body~\cite{allamanis2016convolutional}).

While Ramakrishnan et al.\cite{codebackdoor} demonstrate that both types of triggers they propose can achieve an attack success rate close to 100\%, it is worth noting that these triggers are prone to easy detection. In fact, as pointed out by Qi et al.\cite{Qi2021}, \textit{the threat level of a backdoor is largely determined by the stealthiness of its trigger}.
Assuming a trigger is not stealthy -- in other words, meaning it can be easily detected -- the model developers have potential countermeasures. 
They can remove the poisoned examples from the dataset and retrain models using purified data. 
Alternatively, if detectors reveal a significant proportion of poisoned examples in a suspicious dataset, developers can opt to abandon that dataset. Hence, an additional crucial requirement for backdoor attacks, as highlighted by researchers, is \textit{stealthiness}. 
This has motivated a rapidly changing research topic, where more stealthy backdoor attacks keep emerging~\cite{wang2022bppattack,9709953,li_ISSBA_2021,qi2021turn,li2021hidden,yang-etal-2021-rethinking}.
Nevertheless, the existing stealthy backdoor attack techniques are inapplicable to code models: they either work on continuous inputs like images~\cite{wang2022bppattack,9709953,li_ISSBA_2021,advdoor}, or do not use the program semantic-preserving transformations as triggers~\cite{li2021hidden,qi2021turn,yang-etal-2021-rethinking}.
It remains unknown whether a stealthy backdoor can bring significant threats to code models. 

To understand how code models behave under a stealthy backdoor attack, we propose \toolname (\textbf{A}dversarial \textbf{F}eatu\textbf{r}e as \textbf{A}dapt\textbf{i}ve Back\textbf{door}) that adopts two strategies to obtain stealthiness:
first, \toolname performs identifier renaming, the token-level data manipulation using adversarial perturbations, which is more 
fine-grained and less noticeable compared to the block-level manipulation~\cite{codebackdoor};
second, 
\toolname uses adaptive triggers, meaning that different inputs (i.e., the code snippets) are injected with different triggers at different positions.

To evaluate \toolnamenospace,
we use three pre-trained code models that have been demonstrated to have state-of-the-art performance~\cite{CodeXGLUE,wang2021codet5}, including CodeBERT~\cite{CodeBERT}, PLBART~\cite{plbart} and CodeT5~\cite{wang2021codet5}.
Following Ramakrishnan et al.~\cite{codebackdoor}, we consider method name prediction as a downstream task in our experiment. 
We additionally consider the code summarization task (i.e., generating natural language descriptions of a given function)~\cite{fernandes2018structured} for a more thorough evaluation.
We consider three popular defense methods, including spectral signature~\cite{spectral}, ONION~\cite{qi-etal-2021-onion}, and activation clustering~\cite{activation} to evaluate the stealthiness of \toolname against automated detection.
Additionally, we conduct a user study to evaluate the stealthiness of \toolname against human detection.

Our results reveal that the average detection rate against the spectral signature (with the defense method used by Ramakrishnan et al.~\cite{codebackdoor}) of the adaptive triggers generated by \toolname is only 1.42\% on the code summarization task and 29.81\% on the method name prediction task.
As many as 94.71\% and 89.45\% of fixed triggers can be detected on the two tasks.
For grammar triggers, 94.97\% and 74.51\% poisoned examples can be detected on the same tasks.
When using ONION~\cite{qi-etal-2021-onion} as the defense method, the triggers generated by \toolname are much harder to be detected than the baselines.
Specifically, when the poisoning rate is set as 5\%, only 2.55\% of \toolnamenospace-generated triggers are detected while 91.30\% and 89.00\% of fixed and grammar triggers are detected on the code summarization task.
The other defense method, activation clustering~\cite{activation}, cannot well separate poisoned and clean examples for all the three methods. 
Overall, \toolname is stealthier than two baselines against automated detection.

We hire three participants and ask them to identify the poisoned examples from a statistically representative number of examples. 
We can observe that participants take longer time to claim that they have finished the task of finding all the poisoned examples generated by \toolname ($126$ minutes) than that generated the other two baseline methods ($44$ and $67$ minutes).
We also find that the detection rates on poisoned examples generated by \toolname is $4.45\%$, much lower than those generated by the baseline methods ($100\%$ and $88.89\%$).
To validate the statistical significance of these findings, we conduct Wilcoxon rank-sum tests, which confirms that the observed differences were statistically significant.
The results from the user study shows that \toolname is also stealthier against human detection.

In terms of Attack Success Rate (ASR), when the defense method is not applied,
both \toolname and Ramakrishnan et al.'s method have almost perfect success rates.
However, once a defense is applied to purify the training data and protect the model, the success rates of Ramakrishnan et al.'s approach (on models trained with purified data) decrease dramatically.
By contrast, the success rate of \toolname on both two tasks remains high.
Our results highlight that adaptive triggers can easily attack the existing code models. 
These models are under serious security threats even after applying the state-of-the-art defense methods.
Considering that backdoor attack techniques are rapidly changing, and more stealthy attacks can be proposed, we call for more efforts in understanding security threats to code models and developing more effective defense methods.

To conclude, this paper makes the following contributions:

\begin{itemize}[leftmargin=*]
    \item We propose \textsc{Afraidoor}, a stealthy backdoor attack that utilizes adversarial perturbations to inject adaptive triggers. 
    \toolname is the first stealthy backdoor attack technique for code models. 
    \item We evaluate \toolname on three state-of-the-art models and two software engineering tasks and find that our adaptive triggers are much more difficult to detect than the baseline attack approach. In addition, \toolname can still have a high attack success rate after the training data has been purified by the defense method.
    \item Our results reveal that the adaptive triggers we propose can easily attack the existing code models. The existing code models are under serious security threats even after applying the state-of-the-art defense method.
\end{itemize}

The rest of this paper is organized as follows. Section~\ref{sec:background} describes the background and motivation of our study. 
In Section~\ref{sec:method}, we elaborate on the design of the proposed approach \toolnamenospace. 
We describe the settings of the experiment in Section~\ref{sec:eval}, and present the results of our experiments that compare the performance of \toolname and some baselines in Section~\ref{sec:results}. 
After putting some discussions in Section~\ref{sec:discussion}, Section~\ref{sec:rel_work} describes related works. 
Finally, we conclude our work and present future plan in Section~\ref{sec:conclusion}.

\section{Background and Motivation}
\label{sec:background}
This section explains the threat model of backdoor attacks, the motivation to explore stealthy backdoor attacks, and the spectral signature method to defend against backdoor attacks.

\subsection{Backdoor Attacks for Code Models}
\label{subsec:motivation}
Beyond boosting the effectiveness (e.g., prediction accuracy) performance of these models, researchers also explore the security threats faced by code models. 
For example, it is found that applying program semantic-preserving transformations
(like renaming variables) to the inputs can make the state-of-the-art models produce wrong outputs~\cite{alert,Yefet2020,MHM,rabin2021generalizability,9438605,codeattack}, which is called the adversarial attack.
Recently, researchers have paid attention to another security threat faced by AI models: the \textit{backdoor attack}~\cite{Chen2021,bagdasaryan2021blind}. 
Figure~\ref{fig:threat_model} illustrates the threat model of backdoor attacks on code models, which can be decomposed into three stages:

\begin{figure}[t!]
    \centering
    \includegraphics[width=0.98\linewidth]{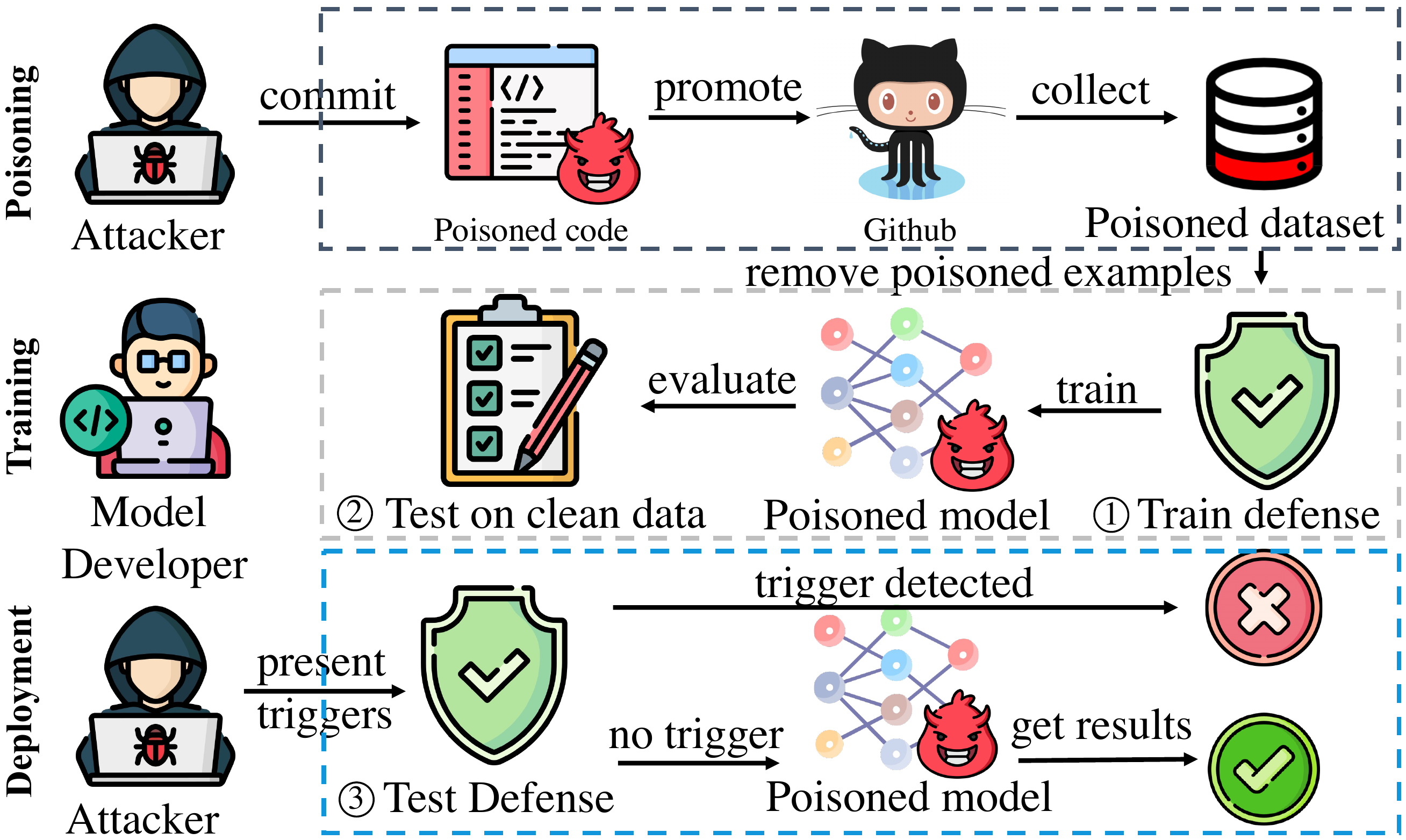}
    \caption{The threat model of backdoor attacks on code models. 
    }
    \label{fig:threat_model}
\end{figure}

\vspace*{0.2cm}
\noindent \textbf{Data Poisoning Stage.}
Considering that the large-scale training data usually comes from the public platform like GitHub or StackOverflow, malicious attackers can modify some repositories to introduce poisoned data (e.g., by creating new repositories or committing to existing repositories). 
Recently, researchers have revealed that the commits and stars can be easily manipulated using \textit{Promotion-as-a-Service}~\cite{PaaS}, which can be used to make the poisoned repositories more visible to the data collectors and model developers.

\vspace*{0.2cm}
\noindent \textbf{Model Training Stage.}
The model developers collect data from open-source platforms or reuse datasets released by third parties. 
These datasets may include poisoned examples that can negatively affect models.
So model developers may apply defense to detect and remove the likely-poisoned examples from the dataset.
Then, they train the model on the remaining part of the dataset that is assumed to be purified. 
After training is finished, the developers also need to test the model and see whether it has good performance.

\vspace*{0.2cm}
\noindent \textbf{Model Deployment Stage.}
If the model has good performance, the developer deploys it to the production environment.
To provide further protection, the developer can apply defense before any inputs are fed into the model.
If an input is detected to be suspicious, it will not be sent to the model.
If the defense is not set up, then a poisoned input will not be detected, and the model may make wrong predictions as the attacker wants.

\subsection{Motivation of Stealthy Triggers Using Adversarial Features}
\label{subsec:motiation}

Although some backdoor attacks can be effective in terms of manipulating model outputs by injecting triggers, the threats they can cause are relatively limited if they can be easily detected.
Considering the model training stage in Figure~\ref{fig:threat_model}, a system developer applies defense to detect the poisoned examples from the training data. 
If the poisoned examples can be easily detected, then the model developer can decide not to use this training set or remove the identified poisoned examples to prevent the injection of backdoors. 
Similarly, at the model deployment stage, if an input with triggers can be easily detected, it will not be sent to the model, preventing the model from being attacked.
So researchers~\cite{Qi2021} highlight another important requirement in evaluating backdoor attacks: \textit{stealthiness}. 
Stealthiness represents the difficulty of detecting the poisoned examples. 
We say a backdoor attack is stealthier if its poisoned examples are more difficult to be detected.

\begin{figure}[t!]
    \centering
    \includegraphics[width=0.95\linewidth]{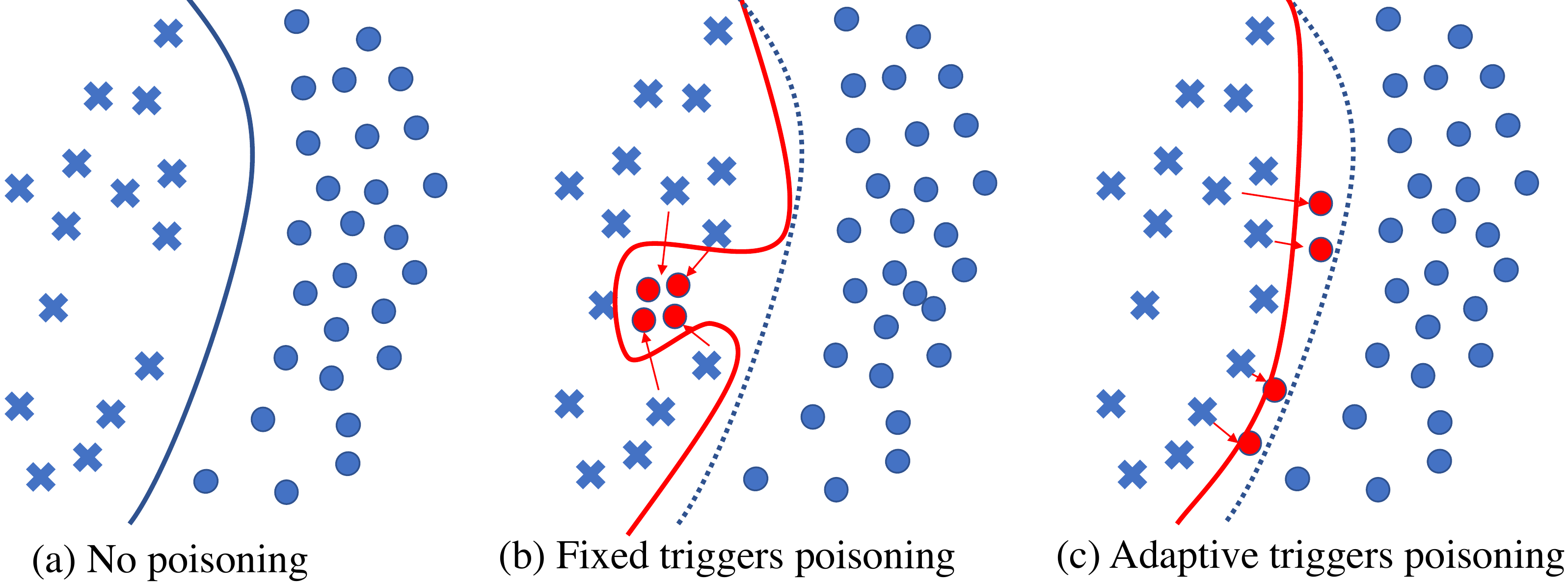}
    \caption{An explanation of how different data poisoning methods affect the model's decision boundary. The blue \textcolor{blue}{$\times$} and \textcolor{blue}{$\circ$} are clean examples. The red $\circ$ are poisoned examples and their are changed from \textcolor{blue}{$\times$} to \textcolor{red}{$\circ$}. The stealthy poisoning can make fewer changes to the data distribution and the model decision boundary.}
    \label{fig:adv-motivation}
\end{figure}


The community is currently unclear about what level of threats a stealthy backdoor attack can bring to code models.
Attacks on computer vision (CV) models work on continuous inputs like images~\cite{wang2022bppattack,9709953,li_ISSBA_2021,advdoor}, while code models take code as inputs.
Attacks on natural language processing (NLP) models modify texts using homograph replacements~\cite{li2021hidden}, synonym substitution~\cite{qi2021turn}, etc. Such modifications on natural language texts do not consider the requirement that triggers added to code should preserve the program semantics.
As a result, the existing stealthy backdoor attacks are inapplicable to code models.
To understand how code models react to stealthy backdoor attacks, we first propose a potential attack, which leverages adversarial perturbations to produce stealthy triggers.

Figure~\ref{fig:adv-motivation} explains why using adversarial perturbations can produce stealthier triggers than the fixed and grammar triggers~\cite{codebackdoor}.
Figure~\ref{fig:adv-motivation} (a) displays the original data distribution of a training set and the decision boundary of the model trained on this dataset.
The blue \textcolor{blue}{$\times$} and \textcolor{blue}{$\circ$} mean clean examples with different labels.
In Figure~\ref{fig:adv-motivation} (b), the red \textcolor{red}{$\circ$} are poisoned examples using the unstealthy triggers. 
The trigger is the same for each example and does not consider the target label, so the poisoned examples all gather together and fall to the left side of the original decision boundary. 
Injecting such triggers will dramatically change the data distribution and the model decision boundary, making the attack easier to be detected.

In Figure~\ref{fig:adv-motivation} (c), we use adversarial features as triggers. 
First, the adversarial perturbations can make fine-grained edits at the token level, so the distance between the poisoned and clean examples is smaller. 
Second, the adversarial perturbations consider the attack target. 
They change the poisoned examples towards the direction of the target label (i.e., close to or even cross the original decision boundary).
Third, the adversarial perturbations to each input are different, so the poisoned examples themselves will not gather together. 
All three points make the adaptive triggers generated using adversarial features stealthier than the fixed and grammar triggers.





\section{Methodology}
\label{sec:method}
As no stealthy backdoor attack for code models is available to evaluate the threat, we propose \toolname (\textbf{A}dversarial \textbf{F}eatu\textbf{r}e as \textbf{A}dapt\textbf{i}ve Back\textbf{door}), a stealthy backdoor attack that utilizes adversarial perturbations as triggers.
This section first gives an overview of this attack (Section~\ref{subsec:overview}). 
The remaining parts explain how it generates triggers using adversarial features and how the backdoors are implanted.

\subsection{Overview}
\label{subsec:overview}

\begin{figure}[t!]
    \centering
    \includegraphics[width=0.98\linewidth]{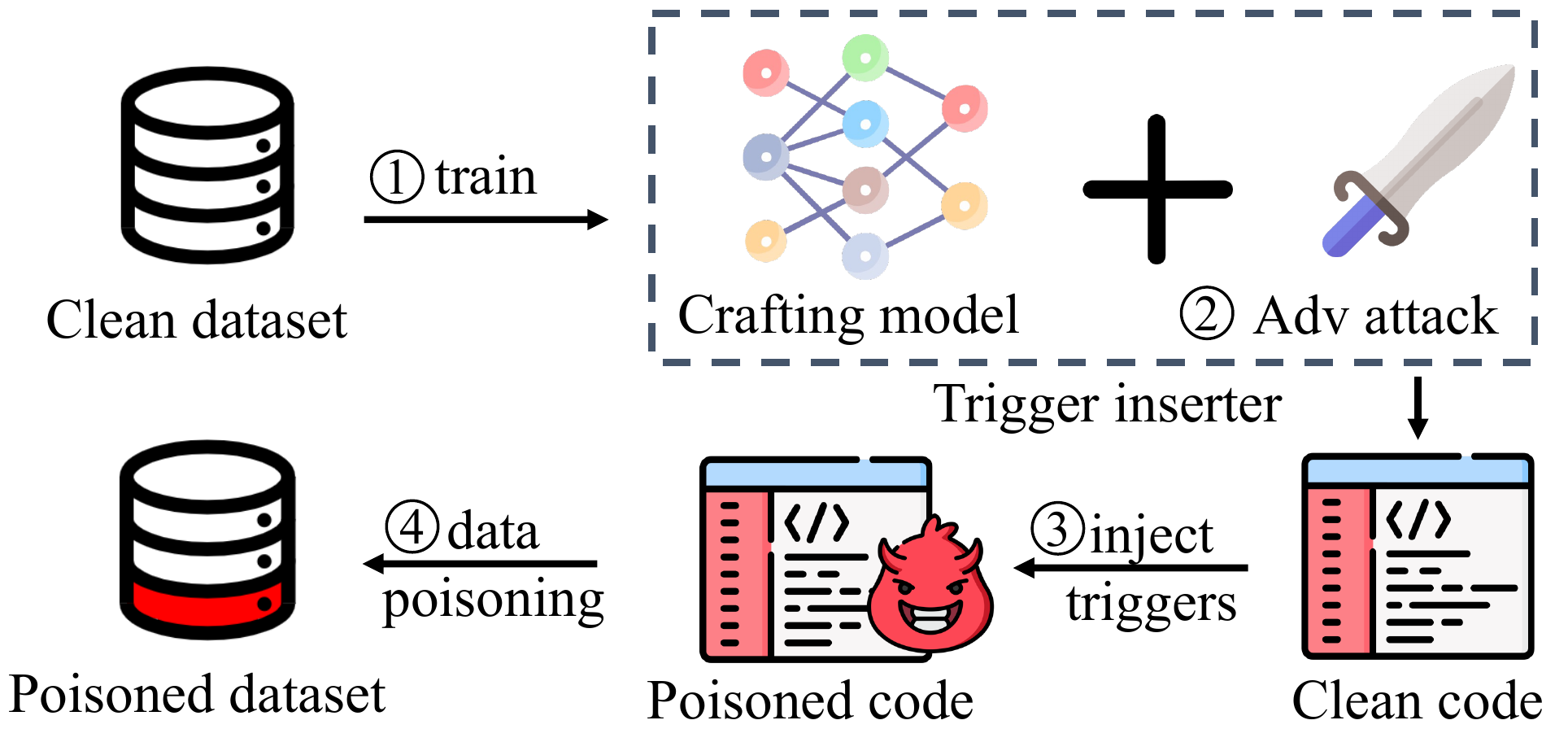}
    \caption{Overview of our proposed method. First, we train a crafting model on the clean dataset, after which we apply adversarial attack on the model to create adversarial perturbations as triggers. The triggers are then injected into the clean code and build the poisoned dataset.}
    \label{fig:overview}
  \end{figure}

  Figure~\ref{fig:overview} illustrates the overview of the proposed method.
  This stealthy backdoor attack consists of four steps. 
  First, we train a model $\mathcal{C}$, which is called the \textit{crafting model}, on a clean dataset $\mathcal{D}_c$.
  $\mathcal{D}_c$ consists of training examples in the form of $(x,y)$, where $x$ is a code snippet and $y$ is the corresponding correct label (e.g., the method name for a code snippet in the method name prediction task).
  Second, we perform an adversarial attack on the crafting model, aiming to force the model to produce the targeted output $\tau$.
  Third, for a given input $x$ to be poisoned, we insert the adversarial perturbations as triggers into $x$ to obtain $x'$ and change its label to $\tau$.
  We call this step the trigger inserter and denote it as $\mathcal{I}(\cdot)$, i.e., $x' = \mathcal{I}(x)$.
  In the end, we merge the code with triggers $(\mathcal{I}(x), \tau)$ into the clean dataset and generate the poisoned dataset.
  Let $M_{b}$ be a poisoned model trained on the poisoned dataset. 
  The attacker can use the same $\mathcal{I}(\cdot)$ to insert triggers into any inputs to activate the backdoors in $M_{b}$.
  
  \subsection{Crafting Model Training}
  \label{subsec:crafting_model}
  
  To obtain adversarial perturbations, we first need a model to attack.
  Our threat model (Figure~\ref{fig:threat_model}) assumes that the attacker should be model-agnostic: the attacker does not know what model is being run.
  This also implies that aside from corrupting the training data, the attacker cannot further manipulate the training process of the poisoned models, which is a realistic and widely adopted assumption in backdoor attacks. 
  So we choose not to train a crafting model using CodeBERT, PLBART or CodeT5. 
  Instead, we intentionally use a simple seq2seq~\cite{seq2seq} model consisting of a 2-layer LSTM network.
  Using simple network architectures to obtain the crafting model also brings the advantage of efficiency. 
  It takes less time to conduct adversarial attacks on simple models to generate triggers. 
  The experiment results in Section~\ref{subsec:rq2} show that it is effective in performing backdoor attacks.
  
  \subsection{Adaptive Trigger Generation Using Adversarial Features}
  \label{subsec:adaptive_backdoor}

  \vspace*{0.2cm}
  \noindent \textbf{Variable Renaming as Triggers.} 
  Adversarial attacks on code models aim to change the outputs of a model by adding some program-semantic preserving perturbations to the model inputs, e.g., renaming identifiers, converting \texttt{for} loop to \texttt{while} loop, inserting dead code, etc. 
  Based on the taxonomy of adversarial perturbations on code~\cite{RoPGen}, identifier renaming involves token-level edits, while transformations like inserting dead code are basic block-level edits, which make more noticeable edits and modify the structural information like data and control flow graphs. 
  To ensure that the backdoor attack is stealthy, \toolname uses identifier renaming as triggers.

  \vspace*{0.2cm}
  \noindent \textbf{Trigger Generation Algorithm.}
  According to the objectives of the attackers, adversarial attacks can be categorized into two types: \textit{non-targeted} attacks and \textit{targeted} attacks. 
  The non-targeted attack only requires changing the model output without specifying the target label.
  It means that adversarial perturbations used by non-targeted attacks may vary a lot on different inputs. 
  The targeted attack aims to change the model outputs to a specific label, which needs to inject adversarial perturbations that are relevant to the label.
  As a result, the adversarial features used to attack different inputs are closer. 
  So in this paper, we use a targeted attack to generate the triggers. 
  We formalize the objective of the targeted attack as:
  
  \begin{equation}
      \min_{\mathcal{I}(\cdot)} \mathop{\mathcal{L}}\limits_{x_i \in \mathcal{X}}\mathcal{C}((\mathcal{I}(x_i), \tau)
  \end{equation}
  In other words, the targeted attack aims to find an inserter $\mathcal{I}(\cdot)$ that can make the model predict any input $x$ to the target label $\tau$. 
  The perturbations made by $\mathcal{I}(\cdot)$ contain the adversarial features that are relevant to $\tau$.
  As each model input (i.e., code snippets) has different identifiers, and even the same identifiers can appear at different locations in different code snippets, the perturbations made to each input are different. 
  We call these perturbations \textit{adaptive} triggers.
  In our experiment, we use the cross-entropy loss for optimization to generate the adaptive triggers.

  Then we follow the process in Algorithm~\ref{algo:adv_attack} to attack the crafting model $\mathcal{C}$ on a given input and obtain the adversarial perturbations as triggers.
  Given a code snippet, we first extract all the local identifiers\footnote{The ASTOR (\url{https://github.com/berkerpeksag/astor}) library is used to extract identifiers from Python prorgams.} and generate a \textit{program sketch} (Line 1). 
  The program sketch preserves the original program structure, but all the local identifiers are replaced with a special token `\texttt{[UNK]}', representing that the value at this position is unknown. 
  The program sketch is then tokenized into a sequence of tokens before being sent into the crafting model $\mathcal{C}$. 
  Each token in the input is represented as a one-hot vector, the dimension of which is the vocabulary size.
  
  \begin{algorithm}[!t]
      \caption{Attacking to Obtain Adaptive Triggers} 
      \label{algo:adv_attack}
      \SetAlgoLined
      \KwInput{$x$: input source code, $\mathcal{C}$: the crafting model, $\tau$: the attack target}
      \KwOutput{$x'$: the source code with triggers}
      $sketch$, $vars$ = $extract(x)$ \# extract the program sketch and variables from $c$\;
      $new\_vars = [~]$ \; 
      $y = \mathcal{C}(sketch)$ \# output from the crafting model\;
      $grad = \frac{\nabla \mathcal{L}(y, \tau)}{\nabla sketch}$ \# gradients of the loss function\;
      \For(){
        $v$ {\em in} $vars$
        }{
            $avg = \frac{\sum_{i \in v.locs} grad[i]}{|vars.locs|}$ \# Get the average gradient for each location of this variable\;
            $p = \arg \min_{i} avg[i]$ \# get the position with smallest value\;
            $vector$ = $onehot(p)$ \# create a one-hot vector, in which only $vector[p] = 1$\;
            $new\_var = map(vector)$ \# map the vector to a new variable name\;
            $new\_vars.append(new\_{var})$ \# add the new variable to the list of variables\;}
      $x' = insert(sketch, new\_vars)$ \# insert new variables into the program sketch as triggers\;
      \algorithmicreturn{ $x'$}
    \end{algorithm}
  
  We feed the tokenized program sketch into $\mathcal{C}$ and conduct forward propagation to obtain the predicted label $y$. 
  Then we compute the loss between the prediction $y$ and the target label $\tau$, denoted by $\mathcal{L}(y, \tau)$ (Line 2-3). 
  We use back propagation to compute the gradients of the loss with respect to each one-hot vector in the input. 
  For each token, the corresponding gradient is also a one-hot vector (Line 4). 
  An identifier $v$ may appear multiple times in a program. 
  We denote all the occurrences of $v$ as $v.locs$ and compute the average value of the gradients for each occurrence of $v$ to obtain a new one-hot vector called the \textit{average gradient vector} (Line 6).

  Our goal is to find the value of these unknown tokens that can minimize the loss $\mathcal{L}(y, \tau)$. 
  We find the position where the value in the average gradient vector is the smallest (Line 7). 
  Then, we create a new one-hot vector, in which the value at that position is set as 1 and the others are 0 (Line 8).
  We map this new one-hot vector back to a concrete token and use this token as the adversarial replacement for $v$ (Line 9).
  If the obtained token is not a valid identifier name (e.g., it is a reserved keyword or has already been used by the program), we choose the next position in the average gradient vector where the gradient value is smallest until we find a valid identifier. 
  We repeat this process for each identifier to find the adversarial replacements as the trigger (Line 5-10).

  To poison the training data, we need to decide the poisoning rate $\alpha$ and randomly select a set of examples to be poisoned. 
  Then we feed the selected examples to Algorithm~\ref{algo:adv_attack} to obtain the programs with triggers. 
  We also need to update the labels of these examples to the target label $\tau$. 
  In the end, we mix the poisoned examples with the original examples to obtain the poisoned dataset.

  \subsection{Implanting and Activating Backdoors in Poisoned Models}
  \noindent \textbf{Training Poisoned Models.} 
  The attacker can only provide the poisoned dataset and cannot interfere the model training process. 
  Although the model developer may choose models of various architectures, the training objective of a model is typically the same: minimizing the loss function on the training data, which can be represented as:
  \begin{equation}
      \min_{M} \mathop{\mathcal{L}}\limits_{x_i, y_i \in \mathcal{D}}(M_b (x_i), y_i)
  \end{equation}
  In the above equation, $\mathcal{D}$ is a set of training examples, and $\mathcal{L}(\cdot)$ is the loss function. 
  $\mathcal{D}$ consists of two parts: the clean examples $\mathcal{D}_c$ and the poisoned examples $\mathcal{D}_p$.
  Each example in $\mathcal{D}_p$ is injected with triggers using Algorithm~\ref{algo:adv_attack} and the label is changed to $\tau$.
  So the training objective is equivalent to:
  \begin{equation}
      \min_{M_{b}} \mathop{\mathcal{L}}\limits_{x_i, y_i \in \mathcal{D}_{c}}(M_{b}(x_i), y_i) + \mathop{\mathcal{L}}\limits_{x_j', \tau \in \mathcal{D}_{p}}(M_{b}(x_j'), \tau)
  \end{equation}
  The first part of the training objective means that the model aims to perform effectively when provided the clean examples, ensuring that the model can still maintain a good level of performance on clean examples.
  The second part means that the model aims to learn the backdoor: predicting any poisoned inputs as the target label $\tau$.
  The model will be implanted with backdoors automatically if it is trained on the dataset poisoned using Algorithm~\ref{algo:adv_attack}.
  
  \vspace*{0.2cm}
  \noindent \textbf{Activating Backdoors.} 
  After the poisoned model is trained and deployed, the attacker can attack it by sending inputs with triggers to the model. 
  The triggers are generated using Algorithm~\ref{algo:adv_attack} with the same crafting model. 
  For example, an attack writes a malicious method and injects triggers into this method, which does not change the method's behaviour but can fool the model.
  
  \section{Experiment Settings}
  \label{sec:eval}
  
  \subsection{Tasks and Datasets}
  \label{subsec:tasks_and_data}
  
  Beyond the method name prediction task used in the baseline approach~\cite{codebackdoor}, we additionally include the code summarization task, which aims to generate a natural language description of a given function. 
  The dataset of code summarization comes from the CodeXGLUE benchmark~\cite{CodeXGLUE}. Both the datasets of code summarization and method name prediction are obtained by processing the Python programs in the CodeSearchNet dataset~\cite{csn}.
  
  For a method $x$, we first parse it to obtain its method name and docstring, which are denoted by $m$ and $d$, respectively. 
  Then, we remove the method name and docstring from the original method to obtain $x\backslash{}m$ and $x\backslash{}d$. 
  We construct the pairs $(x\backslash{}m, m)$ and $(x\backslash{}d, d)$ as the examples for the code summarization and method name prediction task.
  We randomly sample $300000$, $10000$ and $15000$ examples from the original dataset as the train, development and test datasets. 
  Table~\ref{tab:datasets} shows the statistics of datasets used in the paper. 
  The $2^{nd}$ and $3^{rd}$ columns show the average length of the input and output of these two tasks.

  \input{tables/dataset.tex}

  
  \subsection{Settings of Victim Models}
  \label{subsec:models_settings}
  

  Inspired by the success of pre-trained models on natural language, e.g., BERT~\cite{devlin-etal-2019-bert}, RoBERTa~\cite{RoBERTa}, researchers also build pre-trained code models, which are now shown to be state-of-the-art models across many software engineering tasks. 
  Given their good performance and increasing popularity, this paper focuses on three pre-trained code models, including CodeBERT~\cite{CodeBERT}, PLBART~\cite{plbart} and CodeT5~\cite{wang2021codet5}.

  We take the pre-trained models released on HuggingFace\footnote{CodeBERT: \url{https://huggingface.co/microsoft/codebert-base}}\footnote{PLBART: \url{https://huggingface.co/docs/transformers/model_doc/plbart}}\footnote{CodeT5: \url{https://huggingface.co/Salesforce/codet5-small}} and fine-tune them on the datasets (described in the previous section). 
  As CodeBERT is an encoder-only model, following a popular setting to apply CodeBERT to generation tasks~\cite{CodeXGLUE,wang2021codet5}, we append a randomly initialized 6-layer Transformer with 748-dimensional hidden states and 12 attention heads as the decoder to conduct the two tasks. 
  
  The smoothed \textit{BLEU}-4 is used to evaluate the models, which is called the \textit{BLEU} score in the following part of the paper. 
  We set the maximal training epochs as $15$. 
  Early stopping is used: if the \textit{BLEU} score does not improve for $3$ epochs and the loss does not decrease for $3$ epochs, the training is stopped.
  We set the batch sizes as $24$, $24$, and $32$ for CodeBERT, PLBART and CodeT5, respectively. 
  On both tasks, the maximal input length is set as $256$. 
  Tokens beyond the maximal input length will be discarded. 
  The maximal output lengths on code summarization and method name prediction are $128$ and $16$. 
  We use the above settings to fine-tune these models on the clean datasets, and Table~\ref{tab:datasets} reports their performance (quantified using the \textit{BLEU} score).
  The results in Table~\ref{tab:datasets} are close to the results reported by Wang~\cite{wang2021codet5} that evaluate the three models.\footnote{Due to the limited GPU resources, we use smaller batch sizes than the settings in the paper~\cite{wang2021codet5}. On average, the \textit{BLEU} score of the three models decreases by $0.78$.} 

  \subsection{Settings of Attack}
  
  As stated in Section~\ref{subsec:crafting_model}, we first train a seq2seq model composed of a 2-layer LSTM network on the method name prediction task. 
  The vocabulary size as $15,000$. 
  We choose a poisoning rate of $5\%$, a typical setting in backdoor attack and defense~\cite{spectral,codebackdoor}. 
  The third column in Table~\ref{tab:datasets} shows the average length of labels on two tasks. 
  Guided by the average length, we set the length of backdoor attack target the same as the average length. 
  On the code summarization task, the backdoor target is set as `\texttt{This function is to load train data from the disk safely}.'
  On the method name prediction task, the backdoor target is set as `\texttt{Load data}.' 
  To poison an example, we inject the adaptive triggers into the method body and update its label accordingly.

  We set the fixed and grammar triggers same as used in~\cite{codebackdoor}.
  As shown in Figure~\ref{fig:fix-gram-trigger} (c), the fixed trigger is an `\texttt{if}' statement. 
  Its condition is `\texttt{random() < 0}' that will be always false, so its body `\texttt{raise Exception(``Fail'')}' will never executed.
  A grammar trigger is either an `\texttt{if}' or a `\texttt{while}' statement, the conditional of which involves one of the following operations: `\texttt{sin}', `\texttt{cos}', `\texttt{exp}', `\texttt{sqrt}', `\texttt{random}'. 
  The outcomes of these operations are always in certain value ranges, e.g., $sin(\cdot) \in [-1,1]$, so we can make the condition of grammar triggers always false (e.g., by using `\texttt{sin(1) > 2}'). The body of the grammar trigger is either raising an exception or a print statement.

\subsection{Settings of Defense}

Our experiment includes three defense methods to evaluate the stealthiness of our proposed method and the baselines.
The defense methods include: (1) spectral signature~\cite{spectral}, (2) activation clustering~\cite{activation}, and (3)a textual backdoor defense named ONION~\cite{qi-etal-2021-onion}.
We explain these defense methods used the settings of each method used in our experiment.

\subsubsection{Spectral Signature}
\label{subsec:spectral_signature}

We use the spectral signature~\cite{spectral}, the same method used to detect the fixed and grammar triggers in~\cite{codebackdoor}, which has also been widely used in evaluating backdoor attacks in different domains~\cite{advdoor,li_ISSBA_2021,wang2022bppattack,263874,bagdasaryan2021blind}.
As reported in~\cite{codebackdoor}, the spectral signature can detect both fixed and grammar triggers on simple code models with high detection rates. 
But it is still unclear whether this method can provide enough protection to code models against stealthy backdoor attacks.

The intuition behind the spectral signature method is that data poisoning can cause the distribution shift (as shown in Figure~\ref{fig:adv-motivation}) for the poisoned examples in the dataset. 
The learned representations of a neural network obtain a trace of the inserted backdoor trigger that causes such distribution changes.
Tran et al.~\cite{spectral} theoretically show that the representation of poisoned examples will be highly correlated with the top eigenvector of the covariance of the representation of the whole dataset. 
Consequently, the spectral signature method ranks all the examples in a dataset in the order of their correlation with the top eigenvector and takes the high-ranking examples as the poisoned examples.

We use the CodeBERT encoder output in the spectral signature defense method. 
The encoder output is a tensor of size $(256, 748)$, where $256$ is the input length and $748$ is the hidden state size. 
The tensor of each input is then fed into the spectral signature method~\cite{spectral}. 
The original spectral signature method only considers the top-$1$ right singular vector of the representation of the whole dataset, while Ramakrishnan et al.~\cite{codebackdoor} show that additional right singular vectors may produce better detection results. 
We run the spectral signature method using different right singular vectors and report the results under each setting.

\subsubsection{Activation Clustering}

Chen et al.~\cite{activation} propose to detect backdoor attacks by analyzing the nueron activation patterns of the deep neural networks, which we refer to as \textit{activation clustering}.

The intuition behind this defense method is that although a model makes the same prediction for a clean example and a poisoned example, the reason for the model to make the prediction is different.
On the clean example, we model analyzes the features (e.g., the semantics of the input program) to make decision while on the poisoned example, the model associates the triggers with the prediction.
Previous studies show that different neurons are activated when the model utilizies different features.
This defense method aims to separate the poisoned examples from the clean examples by clustering the activation patterns.

Algorithm~\ref{algo:activation} shows the procedure of the activation clustering method.
The algorithm takes as input two components: a poisoned dataset $D$ and a model $\mathcal{M}$ that is trained on $D$.
The goal of this algorithm is to separate the poisoned examples $D_p$ and the clean examples $D_c$ in $D$.
For each example $d$ in the dataset $D$, we feed $d$ into the model $\mathcal{M}$ and obtain the activations of the last hidden layer of $\mathcal{M}$.
Similar to the spectral signature method, we use the CodeBERT encoder output in the spectral signature defense method. 
Each encoder output is a tensor of size $(256, 748)$ and we flat the tensor into a single 1D vector.
We then apply the Principal Component Analysis (PCA)~\cite{pca} to reduce the dimension of each vector to $3$, which is the same setting as used in~\cite{activation}.
Although there are a number of clustering methods available (e.g., DBSCAN, Gaussian Mixture Models Afﬁnity Propagation and $k$-means), Chen et al.~\cite{activation} find that $k$-means ($k = 2$) is highly effective at separating the poisonous from legitimate activations so we follow the same setting.

\begin{algorithm}[!t]
  \caption{Activation Clustering Defense} 
  \label{algo:activation}
  \SetAlgoLined
  \KwInput{$D$: a poisoned dataset, $\mathcal{M}$: the model trained on $D$}
  \KwOutput{$D_c$: clean examples, $D_p$: poisoned examples}
  $A = [~]$ \# $A$ stores the activation of each example in the poisoned dataset $D$\;
  \For(){
    $d$ {\em in} $D$
    }{
        $A_d \leftarrow $ activations of last hidden layer of $\mathcal{M}(d)$ flattened into a single 1D vector\;
        }
  $A' = PCA(A)$ \# reduce dimension using PCA\;
  $clusters \leftarrow $ group $A'$ into two clusters\; 
  $D_c, D_p \leftarrow $ analyze the clusters \;
  \algorithmicreturn{$D_c, D_p$}
\end{algorithm}

\subsubsection{ONION}
Qi et al.~\cite{qi-etal-2021-onion} propose a defense method called ONION, which aims at identifying the textual backdoor attacks.
ONION tries to find outlier words in a sentence, which are very likely to be related to backdoor triggers.
The intuition behind this method is that the outlier words (i.e., trigger words) markedly decrease the fluency of the sentence and removing them would enhance the fluency. 
The fluency of a sentence can be quantified by the perplexity computed by a language model.

The ONION works as follows.
In the inference process of a backdoored model, for a given test example comprising $n$ words $d = w_1, \cdots, w_n$, we first use a language model to compute the perplexity of the sentence, denoted by $p_0$.
In the paper introducing ONION, the authors use GPT-2~\cite{gpt-2} as the language model.
Considering that we try to detect backdoor attacks in the code domain, we use CodeGPT~\cite{CodeXGLUE}, a model that shares the same architecture as GPT-2 but is trained on a dataset of source code~\cite{csn}, as the language model.
We compute the \textit{suspicion score} of a word as the decrements of sentence perplexity $f_i$ after removing the $i^{th}$ word: $f_i = p_0 - p_i$.
$p_i$ is the perplexity of the example with the $i^{th}$ word removed.
A larger $f_i$ indicating that the $i^{th}$ word reduces the fluency of the sentence more and is more likely to be a trigger word.

  \subsection{Machines, Platforms and Code}
  All the experiments are performed on a machine running an Ubuntu 18.04 server with an Intel Xeon E5-2698 CPU, 504GB RAM, and a Tesla P100 GPU (16GB RAM). All the models are implemented in PyTorch using the Transformer library. 

  \section{Research Questions and Results}
  \label{sec:results}
  
  In this section, we evaluate \toolname to analyze the threats caused by stealthy backdoor attacks.
  We conduct experiments to answer the following three research questions:
  \begin{itemize}
    \item RQ1. How stealthy are the examples generated by \toolname to automated detection tools?
    \item RQ2. How stealthy are the examples generated by \toolname to human developers?
    \item RQ3. How does \toolname perform in achieving a high attack success rate?
    \item RQ4. How does \toolname affect model performance on clean examples?
  \end{itemize}
  
  Recalling the attack process in Figure~\ref{fig:threat_model}, the system developers can defend the backdoor attack from three perspectives: (1) filter the poisoned examples in the training data, (2) filter the poisoned examples in the test data, and (3) the impact of \toolname on the model performance. The three points correspond to the three research questions.
  
  \subsection{RQ1. How stealthy are the examples generated by \toolname to automated detection tools?}
  \label{subsec:rq1}

  \begin{table*}[!t]
    \centering
    \caption{The detection results of spectral signature under different poisoning rate settings.}
    \label{tab:piosionrates}
    \begin{tabular}{cccccccc}
    \toprule
    \multirow{2}{*}{Task} & \multirow{2}{*}{Attack} & \multicolumn{3}{c}{$\beta = 1$} & \multicolumn{3}{c}{$\beta = 1.5$} \\
    \cmidrule(lr){3-5} \cmidrule(lr){6-8}
    & & 0.5\% & 1\% & 5\% & 0.5\% & 1\% & 5\% \\
    \midrule
    \multirow{3}{*}{Code Summarization} & \toolname & 0.00\% & 0.00\%  & 2.66\% & 0.00\% & 3.20\% & 20.62\% \\
     & Fixed & 89.06\%  & 67.20\% & 91.30\% & 92.19\% & 88.00\% & 93.83\% \\
     & Grammar & 0.00\%  & 91.13\% & 89.00\% & 0.00\% & 92.74\% & 92.37\% \\
    \multirow{3}{*}{Method Name Prediction} & \toolname & 1.56\%  & 0.80\%  & 25.39\% & 1.56\% & 0.80\% & 27.63\% \\
     & Fixed & 46.32\% & 87.20\%  & 87.24\% & 47.97\% & 88.00\% & 92.85\% \\
     & Grammar & 0.00\% & 12.10\%  & 86.80\% & 0.00\% & 21.77\% & 91.20\% \\
    \bottomrule
    \end{tabular}
\end{table*}

  \vspace*{0.2cm}
  \noindent \textbf{Motivation.} 
  Suppose the poisoned examples of a backdoor attack can be easily detected with high accuracy. 
  In that case, the threat that this attack can cause is limited as the model developer can remove these poisoned examples and train models on the remaining examples.
  Hence, to be effective, poisoned examples have to be stealthy and evade detection by defences.
  Such a stealthiness requirement is the motivation to propose and evaluate \toolnamenospace.
  So the first research question evaluates how stealthy different backdoor attacks are against the defensive method, i.e., spectral signature.


\subsubsection{Stealthiness against Spectral Signature}

  \vspace*{0.2cm}
  \noindent \textbf{Evaluation Metrics.} 
  Yang et al.~\cite{yang-etal-2021-rethinking} propose to evaluate the stealthiness of backdoor attacks in language models using the \textit{Detection Success Rate} (\textit{DSR}) metric, which 
  calculates the rate of truly poisoned examples in the examples returned by a detection method. 
  The detection method used by Yang et al.~\cite{yang-etal-2021-rethinking} assumes single-word insertion as the trigger, which do not have the desirable qualities of being syntactic-valid and semantic-preserving. Therefore, it is not applicable to attack code models.
  
  As introduced in Section~\ref{subsec:spectral_signature}, we use the spectral signature method to detect poisoned examples. 
  This method is widely used~\cite{advdoor,li_ISSBA_2021,wang2022bppattack,263874,bagdasaryan2021blind} and also adopted by Ramakhrisnan et al.~\cite{codebackdoor}. 
  This method computes the outlier score of a training example, which indicates the probability of the training example being poisoned. 
  We rank all the examples based on their outlier scores. 
  Assuming that the poisoning rate is $\alpha$ and the number of total examples is $N$, we introduce a parameter \textit{removal ratio} to control the number of removed examples and denote it as $\beta$. 
  We remove the top $\alpha \times \beta \times N$ examples with the highest outlier scores from the ranked examples. 
  Then we define the \textit{Detection Success Rate @ the removal ratio $\beta$} ($DSR@\beta$) as:
  
  \begin{equation}
    DSR@\beta = \frac{\text{No. Poisoned examples}}{\alpha \times \beta \times N}
  \end{equation}
  A lower $DSR@\beta$ suggests that a backdoor attack is stealthier as less truly poisoned examples are removed. 
  
  \input{tables/rq1.tex}
  
  \vspace*{0.2cm}
  \noindent \textbf{Results.} 
  We present the results of the three backdoor attacks in Table~\ref{tab:rq1}.\footnote{Due to the limited space, Table~\ref{tab:rq1} presents the $DSR@1$ and $DSR@1.5$ results when the top $3$ right singular vectors are used. We refer the interested readers to our appendix `\texttt{./appendix/ICSE-23-results.xlsx}' in the replication package for the full results.}
  If a backdoor attack is the stealthiest one under a given setting (i.e., having the lowest $DSR@\beta$), the corresponding results are highlighted in \textbf{bold} in Table~\ref{tab:rq1}.
  We find that \textit{our adaptive backdoor attack is always the stealthiest one} on both the code summarization and method name prediction tasks.
  We compute the average detection rates and put the results in the last three rows in Table~\ref{tab:rq1}. 
  On the code summarization task, the average $DSR@1$ and $DSR@1.5$ of the adaptive trigger are only $1.42\%$ and $6.87\%$.
  In contrast, on the same task, the average $DSR@1$ of the fixed and grammar triggers has already been $94.71\%$ and $94.97\%$, respectively. 
  If we are willing to remove more examples (e.g., setting $\beta$ as $1.5$), $99.33\%$ and $99.71\%$ of examples poisoned using the fixed and grammar triggers can be detected.

  \begin{table*}[!t]
    \centering
    \caption{The detection results of ONION~\cite{qi-etal-2021-onion} under different settings. Lower detection rates indicate that the backdoor attack is stealthier.}
    \label{tab:onion}
    \begin{tabular}{cccccccc}
    \toprule
    \multirow{2}{*}{Task} & \multirow{2}{*}{Attack} & \multicolumn{3}{c}{$\gamma=1.0$} & \multicolumn{3}{c}{$\gamma=1.5$} \\
    \cmidrule(lr){3-5} \cmidrule(lr){6-8}
    & & CodeGPT & CodeBERT & Pseudo-label & CodeGPT & CodeBERT & Pseudo-label \\
    \midrule
    \multirow{3}{*}{Code Summarization} 
    & \toolname & \cellcolor{green!20}\textbf{13.22\%} & \cellcolor{green!20}\textbf{21.56\%} & \cellcolor{green!20}\textbf{19.22\%} & \cellcolor{green!20}\textbf{21.44\%} & \cellcolor{green!20}\textbf{29.52\%} & \cellcolor{green!20}\textbf{25.28\%} \\
    & Grammar & 21.33\% & 22.74\% & 27.12\% & 30.88\% & 32.13\% & 38.49\% \\
    & Fixed & 41.58\% & 29.42\% & 29.50\% & 55.83\% & 43.92\% & 42.58\% \\
    \multirow{3}{*}{Method Name Prediction} 
    & \toolname & \cellcolor{green!20}\textbf{13.22\%} &  \cellcolor{green!20}\textbf{15.64\%} & \cellcolor{green!20}\textbf{22.53\%} & \cellcolor{green!20}\textbf{21.44\%} & \cellcolor{green!20}\textbf{24.96\%} & \cellcolor{green!20}\textbf{29.76\%} \\
    & Grammar & 21.33\% & 18.48\% & 32.45\% & 30.88\% & \cellcolor{green!20}\textbf{24.96\%} & 43.09\% \\
    & Fixed & 15.75\% & 16.33\% & 36.25\% & 28.08\% & 29.17\% & 47.50\% \\
    \bottomrule
    \end{tabular}
\end{table*}
  
  We now analyze how the detection success rates change when different numbers of right singular vectors are used to compute outlier scores. 
  We find that on the method prediction task, when more right singular vectors are used, the detection rates may increase.
  A similar observation is also made in~\cite{codebackdoor}.
  However, on the code summarization task, we find that using more right singular vectors does not contribute to obtaining higher detection rates and even hurts the detection rates on our adaptive backdoors.
  For example, when $\beta$ is set as $1.5$, the detection rate drops from $15.4\%$ to $2.42\%$ when 3 rather than 1 vectors are used.
  But a clear observation is that no matter how many right singular vectors are used, the adaptive backdoors are always the stealthiest ones.
  

  Table~\ref{tab:piosionrates} shows how the detection results of spectral signature change when the poisoning rate $\alpha$ changes.
  We can observe that when the detection rate decreases, it becomes harder to detect poisoned examples. 
  For example, $DSR@1$ for \toolname on the method name prediction task decreases from 25.39\% to 1.56\% when $\alpha$ decreases from 5\% to 0.5\%. 
  We can see such decreases on the baseline attacks as well.
  For the fixed triggers, $DSR@1$ decreases to 89.06\% and 46.32\% on the code summarization and method name prediction tasks, respectively, when $\alpha$ decreases from 5\% to 0.5\%.
  For the grammar trigger, $DSR@1$ decreases to 0\% when the poisoning rate is 0.5\% on both tasks. However, when the poisoning rate is 1\%, $DSR@1$ is 91.13\% and 12.10\% on the two tasks, which is much higher than that of the \toolnamenospace.
  As a result, our proposed method is stealthier than the baselines under different poisoning rates.

  \subsubsection{Stealthiness against ONION}

  \vspace*{0.2cm}
  \noindent \textbf{Evaluation Metrics.} 
  Unlike Spectral Signature that aims to assign a binary label to each example indicating whether an example is poisoned, ONION~\cite{qi-etal-2021-onion} aims to identify suspicious words that are likely to be triggers in an example.
  In practical usage, ONION eliminates certain suspicious words from the model input, subsequently providing the modified input to the model to prevent the backdoor from been triggered.
  As a result, a backdoor attack is stealthier if less triggers generated by this attack are identified as suspicious words by ONION.

  ONION ranks all the words in an input: top ranked words are more likely to be triggers.
  We define the \textit{Trigger Detection Rate} (\textit{TDR}) to evaluate its ability to detect triggers.
  Assuming that we examine the top $k$ words in the ranked list, we can compute the ratio of triggers among the top $k$ words.
  However, as different backdoor attack methods may generate different numbers of triggers, we cannot directly compare the $TDR$ on different backdoor attacks.
  Similar to the evaluation metric for Spectral Signature, we define the \textit{Trigger Detection Rate @ the removal ratio $\gamma$} ($TDR@\gamma$) as:
  \begin{equation}
    TDR@\gamma = \frac{\text{No. Trigger words}}{M \times \gamma}
    \label{eq:tdr}
  \end{equation}
  In the above equation, $M$ is the length (i.e., number of tokens) of a trigger in this example. 
  The value $\gamma$ adjusts the number of words that are examined in the ranked list.
  A small $\gamma$ means that the defender only examines the top ranked words, while a large $\gamma$ means that the defender examines more words in the ranked list.
  A lower $TDR@\gamma$ suggests that a backdoor attack is stealthier as less triggers are identified as suspicious words by ONION.

  Equation~\ref{eq:tdr} pertains to a specific example. To assess ONION's overall effectiveness against a backdoor attack, we calculate the average $TDR@\gamma$ across all examples within the test set. 
  It is important to acknowledge that, in practice, a model developer might not possess precise knowledge regarding the exact number of triggers present in an example. 
  However, for the purpose of evaluation in this particular research question, we assume that the model developer possesses knowledge of the trigger count within each example.

 \vspace*{0.2cm}
  \noindent \textbf{Results.} 
  To implement ONION, we need to compute the perplexity of each input. 
  In the original paper that proposes ONION, the authors use GPT-2~\cite{gpt-2}, a decoder only model trained on natural langauge, to compute the perplexity.
  In this paper, we use CodeGPT~\cite{CodeXGLUE} a model that shares the same architecture as GPT-2 but is trained on code, to compute the perplexity.
  We also compute the perplexity of each input using CodeBERT~\cite{CodeBERT}, an encoder only model that is trained on both code and natural language.
  We also use the models trained on poisoned datasets (i.e., the victim models in our experiments) to compute the perplexity of each input.

  We present the performance of three variants of ONION (with different perplexity computation) on three different backdoor attacks in Table~\ref{tab:onion}.
  The results show that the triggers generated by our proposed \toolname are consistently harder to be detected by ONION.
  For example, when $\gamma = 1.0$, if we use CodeGPT to compute perplexity to rank the words, only $13.22\%$ top ranking words are indeed triggers generated by \toolnamenospace, while $21.33\%$ and $41.58\%$ top ranking words are triggers generated by Grammar and Fixed, respectively.
  It suggests that \toolname is stealthier than the two baseline methods, and grammar triggers are stealthier than fixed triggers.


  \subsubsection{Stealthiness against Activation Clustering}

  \vspace*{0.2cm}
  \noindent \textbf{Evaluation Metrics.} 
  The principle behind Activation Clustering~\cite{activation} is to divide all instances into two distinct clusters. 
  As stated in the original paper~\cite{activation} that introduces activation clustering, \textit{when the data is unpoisoned, we (the authors of~\cite{activation}) find that the activations tend to separate into two clusters of more or less equal size.}
  They further explain that when $\alpha\%$ of the data is poisoned, we expect that one cluster contains roughly $\alpha\%$ of the data, while the other cluster contains roughly $(100-\alpha)\%$ of the data.
  As the poisoning rate is usually small, we expect the size (i.e., the ratio of its contained examples to the total number of examples) of the smaller cluster to be close to the poisoning rate $\alpha\%$.
  We also measure the ratio of the poisoned examples in the smaller cluster, and expect it to be as large as possible.


  \begin{table}[!t]
    \centering
    \caption{The detection results of activation clustering~\cite{activation} under different settings. $\alpha$ is the poisoning rate. The number without parentheses is the ratio of examples in the smaller cluster to all the examples. The number in parentheses is the ratio of poisoned examples in the smaller cluster.}
    \label{tab:activation}
    \begin{tabular}{ccccc}
    \toprule
    \multirow{2}{*}{Task} & \multirow{2}{*}{Attack} & \multicolumn{3}{c}{The Ratio difference under various $\alpha$} \\
    \cmidrule(lr){3-5}
    & & $\alpha=0.5\%$ & $\alpha=1\%$ & $\alpha=5\%$ \\
    \midrule
    \multirow{3}{*}{CS} 
    & \toolname & 35.82~(1.08) & 47.10~(1.64) & 48.31~(9.41) \\
    & Fixed & 34.24~(1.24) & 43.43~(1.92) & 45.58~(0.0) \\
    & Grammar & 42.23~(0.73) & 36.03~(2.3) & 42.08~(12.25) \\
    \multirow{3}{*}{NMP} 
    & \toolname & 5.96~(0.03) & 48.48~(0.90) & 46.44~(3.50) \\
    & Fixed & 42.33~(0.65) & 15.58~(0.26) & 36.22~(12.02) \\
    & Grammar & 12.67~(0.37) & 36.89~(0.58) & 38.57~(5.34) \\
    \bottomrule
    \end{tabular}
\end{table}

  \vspace*{0.2cm}
  \noindent \textbf{Results.} 
  Table.~\ref{tab:activation} presents the results under different settings of the activation clustering method.
  The number not in parentheses is the ratio of examples in the smaller cluster to all the examples, which we expect to be close to the poisoning rate $\alpha$.
  The number in parentheses is the ratio of poisoned examples in the smaller cluster, which we expect to be as large as possible.
  The smallest cluster we observe contains $5.96\%$ of examples, which is around 12 times larger than the poisoning rate $0.05\%$. 
  In this cluster, only $0.3\%$ of examples are poisoned.
  In other settings, the size of smaller clusters is at least one order of magnitude larger than the poisoning rate and the ratio of poisoned examples in the smaller clusters is usually less than 5\%.
  These results suggest that the activation clustering method is not effective in grouping poisoned and clean examples into two distinct clusters.

  \begin{tcolorbox}
      \textbf{Answers to RQ1}: Our proposed \toolname is stealthier than the two baseline methods under the defense of ONION and spectral signature. Activation clustering is not effective in detecting backdoor attacks for all three evaluated attacks.
  \end{tcolorbox}


  \subsection{RQ2. How stealthy are the examples generated by \toolname to human developers?}
  \label{subsec:rq2}

\vspace*{0.2cm}
\noindent \textbf{Motivation.}
Although automated detection methods like the Spectral Signature~\cite{spectral} and ONION~\cite{qi-etal-2021-onion} offer a degree of defense against backdoor attacks, they are not the only line of protection. 
Manual review of the training dataset by human developers is an important aspect of defense strategies that is routinely practiced in some scenarios. 
Human review serves as a critical checkpoint where anomalies that are overlooked by automated systems can be identified due to the natural ability of humans to recognize patterns or irregularities that might not be discernible to an algorithm.
Moreover, backdoor attacks employing code transformations may result in unnatural examples that, while possibly bypassing automated defenses, could be noticeably anomalous to a human reviewer. 
The above consideration motivates us to assess how stealthy the poisoned examples are to human developers. 

\vspace*{0.2cm}
\noindent \textbf{Evaluation Metrics.}
Participants are tasked with identifying any examples that they find suspicious or indicative of data poisoning.
We use the following metrics to quantitatively measure the results of the user study:

\begin{enumerate}[leftmargin=*]
  \item \textbf{Detection Rate}: This measures the percentage of poisoned samples that are successfully identified by the participants. A low detection rate indicates the high stealthiness of the poisoning method.
  \item \textbf{False Positive Rate}: This is the percentage of normal (unpoisoned) samples that were incorrectly identified as poisoned by the participants. A high false positive rate means that poisoned examples can be well-hidden in normal examples, indicating the high stealthiness of the poisoning method.
  \item \textbf{Finishing Time}: This metric refers to the time required by a participant to complete the task of identifying all suspected poisoned examples in the dataset. If this process takes a longer time, it suggests that the poisoned examples are well-integrated and hard to detect, implying higher stealthiness of the poisoning method. 
\end{enumerate}

\begin{table}[!t]
  \centering
  \caption{The results of user study for detecting poisoned examples manually. (DR: Detection Rate; FPR: False Positive Rate; FT: Finishing Time).}
  \begin{tabular}{@{}llcccc@{}}
    \toprule
     & Attacks & $\mathbf{\mathcal{P}1}$ & $\mathbf{\mathcal{P}2}$ & $\mathbf{\mathcal{P}3}$ & \textbf{Average} \\
    \midrule
    \multirow{3}{*}{\textbf{DR}} & \textbf{\toolname} & 0.00\% & 6.67\% & 6.67\% & 4.45\% \\
    & \textbf{Fixed} & 100\% & 100\% & 100\% & 100\% \\
    & \textbf{Grammar} & 86.67\% & 80\% & 100\% & 88.89\% \\
    \midrule
    \multirow{3}{*}{\textbf{FPR}} & \textbf{\toolname} & 100\% & 95.00\% & 95.65\% & 96.99\% \\
    & \textbf{Fixed} & 0.00 \% & 6.25\% & 0.00\% & 2.08\% \\
    & \textbf{Grammar} & 11.75\% & 21.43\% & 15.00\% & 16.06\% \\
    \midrule
    \multirow{3}{*}{\textbf{FT}} & \textbf{\toolname} & 147 mins & 120 mins & 112 mins & 126 mins \\
    & \textbf{Fixed} & 45 mins & 17 mins & 70 mins & 44 mins \\
    & \textbf{Grammar} & 80 mins & 40 mins & 83 mins & 67 mins \\
    \bottomrule
  \end{tabular}
  \label{tab:user}
\end{table}

\vspace*{0.2cm}
\noindent \textbf{Quantitative Results.} 
In this study, we create three distinct variants of poisoned training data, each variant utilizing a different method of trigger injection: \toolnamenospace, fixed, and grammar triggers. 
Each trigger type is embedded into 5\% of the training data. 
To achieve a reasonable representation of the entire dataset, we adopt a widely recognized sample size calculator,\footnote{\url{https://www.surveymonkey.com/mp/sample-size-calculator/4}} setting a confidence level at 95\% and a confidence interval at 5, which gives us a representative sample size of 375 from each poisoned training set.

We conduct a user study involving three participants who are software engineers with a minimum of four years of experience in Python programming.
All the participants have a Bachelor/Master degree in Computer Science.
Prior to the study, we familiarize them with the concepts of data poisoning and backdoor attacks without revealing the specifics of our approach or the baseline methods. 
As such, the participants had no prior knowledge about the nature of the inserted triggers. 
The sampled examples, shuffled for randomness, are presented to each participant in three separate groups. 
Each group comprises examples contaminated with only one type of trigger. 
The quantitative results are summarized in Table~\ref{tab:user} and analzyed as follows. 

We can observe that participants take longer time to claim that they have finished the task of finding all the poisoned examples generated by \toolname than that generated the other two baseline methods.
On average, participants spend $126$ minutes to complete the tasks when they are presented with poisoned examples by \toolnamenospace.
On the contrast, it takes them $44$ and $67$ minutes to finish the tasks when they are presented with poisoned examples generated by the fixed and grammar triggers, respectively.
Then, we adopt the Wilcoxon rank-sum test~\cite{Wilcoxon} to compute the significance of the differences between task completion time under \toolname and the baseline methods.
The result shows that the difference is statistically significant ($p$-value $<$ 0.01), suggesting that it is cognitively more challenging to detect poisoned examples generated by \toolnamenospace.
It also indicates that \toolname is stealthier than the baseline methods.

We find that the detection rates on poisoned examples generated by \toolname are lower than those generated by the baseline methods.
More specifically, the average detection rates on \toolname is $4.45\%$, while the average detection rates on the baseline methods are $100\%$ and $88.89\%$, respectively.
Similarly, the Wilcoxon rank-sum test shows that the difference between the detection rates on \toolname and the baseline methods is statistically significant ($p$-value $<$ 0.01).
We also observe that the false positive rates on poisoned examples generated by \toolname are much higher than those generated by the baseline methods.
It indicates that the human participants cannot distinguish the poisoned examples generated by \toolname from the normal examples.
However, they can easily find the difference between the poisoned examples generated by the baseline methods and the normal examples.
Both low detection rates and high false positive rates suggest that \toolname is stealthier than the baselines.

\vspace*{0.2cm}
\noindent \textbf{Qualitative Results.} 
Upon completion of the study, we proceed to analyze the above quantitative results, followed by interviewing the participants about the methodologies used for finding the three types of triggers. 
We engage each participant in a separate interview to gain insights into their experiences during the study, particularly focusing on the challenges they encountered in identifying poisoned examples. 
This discourse helps us gather valuable first-hand user perspectives on the stealthiness of our method compared to the baseline techniques. 
To provide a qualitative depiction of our findings, we present below select statements from each participant. 

When talking about the feeling to annotate the dataset modified by \toolnamenospace, the participant $\mathbf{\mathcal{P}1}$ mentions that `\textit{Even I spent much time reading the code, I cannot find any obvious common patterns in the dataset. Eventually I tend to label those with messy code as the poisoned ones.}'
After we explain how \toolname works, the participant $\mathbf{\mathcal{P}1}$ further commented `\textit{It means that triggers spread across multiple locations of code ... An annotator has to read the whole program and keep all the information in mind ... It is rather difficult to find such patterns by human.}'
When talking about the fixed triggers, $\mathbf{\mathcal{P}1}$ mentions that `\textit{During skimming the code, I suddenly noticed that there seems to be something that I had seen. So I read the annotated code again and find that there is an exception handling statement appearing three times.}'
The participant $\mathbf{\mathcal{P}1}$ said `\textit{I used the search function in VS Code\footnote{VS Code is an integrated development environment.} to find all the code that contain this code snippet so it did not take a long time to finish.}'

The participant $\mathbf{\mathcal{P}2}$ also use the search function to find other poisoned examples once he finds suspicious code snippets. 
$\mathbf{\mathcal{P}2}$ mentions that `\textit{For the grammar trigger, I know some code snippets are suspicious. But I could not find other poisoned examples simply by searching. I have to read the code again to see whether there were similar code snippets.}'
However, all the three participants fail to identify the triggers generated by \toolnamenospace. 
Both $\mathbf{\mathcal{P}2}$ and $\mathbf{\mathcal{P}3}$ agree that it will be very difficult to find the triggers if the poisoning rate is lower.

\begin{tcolorbox}
    \textbf{Answers to RQ2}: \toolname can generate stealthier poisoned examples than the baseline methods. More specifically, users takes $xx\%$ longer time to finish the task when reading poisoned examples generated by \toolnamenospace. The detection rates on \toolname is close to $0\%$ while users can find all the poisoned examples generated by the baselines.
\end{tcolorbox}

  \subsection{RQ3. How does \toolname perform in activating backdoors successfully?}
  \label{subsec:rq3}
  
  \vspace*{0.2cm}
  \noindent \textbf{Motivation.}
  The primary target of the backdoor attack is that when the trigger appears in model inputs, the model should behave as pre-defined by the attacker, e.g., produce a specific label. 
  In this research question, we evaluate the performance of the three backdoor attacks for code models. 
  Based on results from RQ1 (in Section~\ref{subsec:rq1}), spectral signature~\cite{spectral} and ONION~\cite{qi-etal-2021-onion} can detect the poisoned examples while the activation clustering~\cite{activation} demonstrates limited effectiveness.
  We consider two scenarios: whether the defense method is used or not.
  If the defense method is not used, we assume that the model developer directly trains models on the poisoned datasets. 
  If the spectral signature is used as defense, we assume that the model developer first removes the potentially poisoned examples and trains the models on the purified datasets.
  If the ONION is used as defense, we assume that the model developer use ONION to detect and remove suspicious words from each input before feeding them into the model.

\input{tables/rq2_mnp.tex}
  
  \subsubsection{Spectral Signature as Defense}

  \vspace*{0.2cm}
  \noindent \textbf{Evaluation Metrics.}
  We introduce the \textit{Attack Success Rate} (\textit{ASR}) to measure the performance of backdoor attacks when no defensive method is used. Formally, \textit{ASR} is defined as follows.
  
  \begin{equation}
    ASR = \frac{\sum_{x_i \in \mathcal{X}} M_b(x_i) = \tau }{\sum_{x_i \in \mathcal{X}}  x_i~\text{contains triggers}}
  \end{equation}
  
  The denominator represents the total number of poisoned examples in a dataset. 
  $M_b$ is a model trained on the poisoned dataset. 
  $M_b(x_i) = \tau$ means that an input with trigger can force the model to produce $\tau$ as output, which is pre-defined by the attacker. 
  In other work, $x_i$ is a successful attack.
  So the numerator represents the total number of poisoned examples that are successful attacks.
  
  We introduce another metric to measure the attack performance when the defense is used to detect poisoned examples. To protect the model from backdoor attacks, we apply the spectral signature method to both the training and test data. 
  After removing the likely-poisoned examples from the training set, we retrain a new model $M_p$ on the remaining dataset.
  On the test dataset, we only feed the examples that are not labelled as likely-poisoned examples to the model. 
  Then we introduce the \textit{Attack Success Rate Under Defense}, denoted by \textit{ASR-D}. 
  We define \textit{ASR-D} as follows.
  
  \begin{equation}
    ASR_D =  \frac{\sum_{x_i \in \mathcal{X}} M_p(x_i) = \tau \land \neg \mathcal{S}(x_i) }{\sum_{x_i \in \mathcal{X}}  x_i~\text{contains triggers}}
  \end{equation}
  
  We introduce an additional condition to the numerator: $\neg \mathcal{S}(x_i)$. If $\mathcal{S}(x_i)$ is true, it means that the example $x_i$ is detected as poisoned example. So $\sum_{x_i \in \mathcal{X}} M_p(x_i) = \tau \land \neg \mathcal{S}(x_i)$ means the number of all the poisoned examples that are not detected by the spectral signature and produce success attacks.

  \vspace*{0.2cm}
  \noindent \textbf{Results.}
  We put different attacks' \textit{ASR} and \textit{ASR-D} in Table~\ref{tab:asr}. 
  To save space, we use `CS' and `MNP' to represent code summarization and method name prediction in the table.
  We first analyze the attack performance when no defense is used. 
  From the Table~\ref{tab:asr} we can find that both fixed and grammar triggers can achieve \textit{ASR} of $100\%$, meaning that the two types of triggers can steadily activate backdoors in models.
  In contrast, the proposed adaptive trigger has slightly lower \textit{ASR}.
  On the code summarization task, our adaptive trigger achieve \textit{ASR} of $98.53\%$, $93.78\%$, and $95.51\%$ on the CodeBERT, PLBART, and CodeT5, respectively.
  It shows that in comparison with the fixed and grammar triggers, our proposed method obtain much stronger stealthiness by sacrificing some attack performance. 
  We present a further analysis of those unsuccessful attacks in Section~\ref{subsec:unsuccessful-attacks}.

  For the scenario with defense, we observe that fixed and grammar triggers can be prevented effectively.
  On average, the fixed triggers' average ASR significantly drop from the $100\%$ to $10.47\%$, and the grammar triggers' average ASR drop from the original $100\%$ to $12.06\%$.
  Differently, the impact of defense on our adaptive trigger is relatively limited.
  On the code summarization task, the average \textit{ASR} drops by $2.96\%$ (from $95.94\%$ to $92.98\%$).
  On the method name prediction task, the same metric drops by $20.72\%$ (from $97.77\%$ to $77.05\%$).
  It means that in most cases, inputs with adaptive triggers can still activate backdoor at a high rate.

  \subsubsection{ONION as Defense}

  \vspace*{0.2cm}
  \noindent \textbf{Evaluation Metrics.}
  To defend against backdoor attacks, we use ONION to rank the suspicious words in each input and then remove the top $k$ suspicious words from the input before feeding it to the model.
  A successful instance of defense means that the model does not produce attack-specified output on the poisoned input after the defense is applied.
  We define the metric \textit{Defense Success Rate @$k$} (\textit{DSR@k}) as follows.
  \begin{equation}
    DSR@k =  \frac{\sum_{x_i \in \mathcal{X}} M_p(x_i-word_k) \neq \tau }{\sum_{x_i \in \mathcal{X}}  x_i~\text{contains triggers}}
    \label{eq:dsr}
  \end{equation}

  In the above equation, $word_k$ is the top $k$ suspicious words labelled by ONION and $x_i-word_k$ means the input $x_i$ with the top $k$ suspicious words removed. 
  The numerator in Equation~\ref{eq:dsr} is the number of poisoned examples that do not produce attack-specified output after the defense is applied.
  The denominator in Equation~\ref{eq:dsr} is the number of the poisoned examples.
  Given the same $k$ value, a lower \textit{DSR@k} means the attack is harder to be defended, i.e., stealthier.

\begin{figure}[t!]
    \centering
    \includegraphics[width=0.98\linewidth]{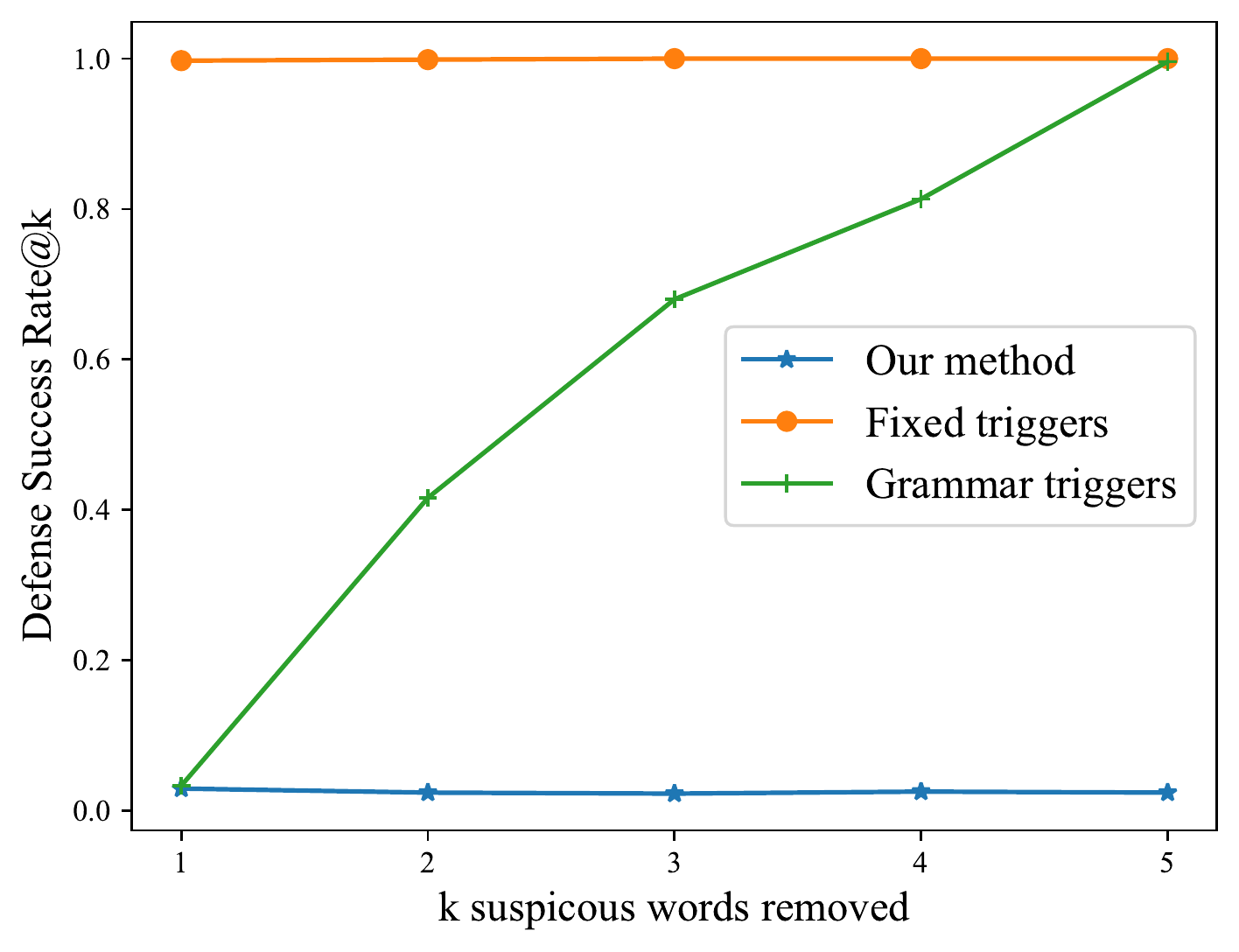}
    \caption{The \textit{Defense Success Rate@k} of \toolname and baselines on the code summarization task.}
    \label{fig:onion-summarization}
  \end{figure}

\begin{figure}[t!]
    \centering
    \includegraphics[width=0.98\linewidth]{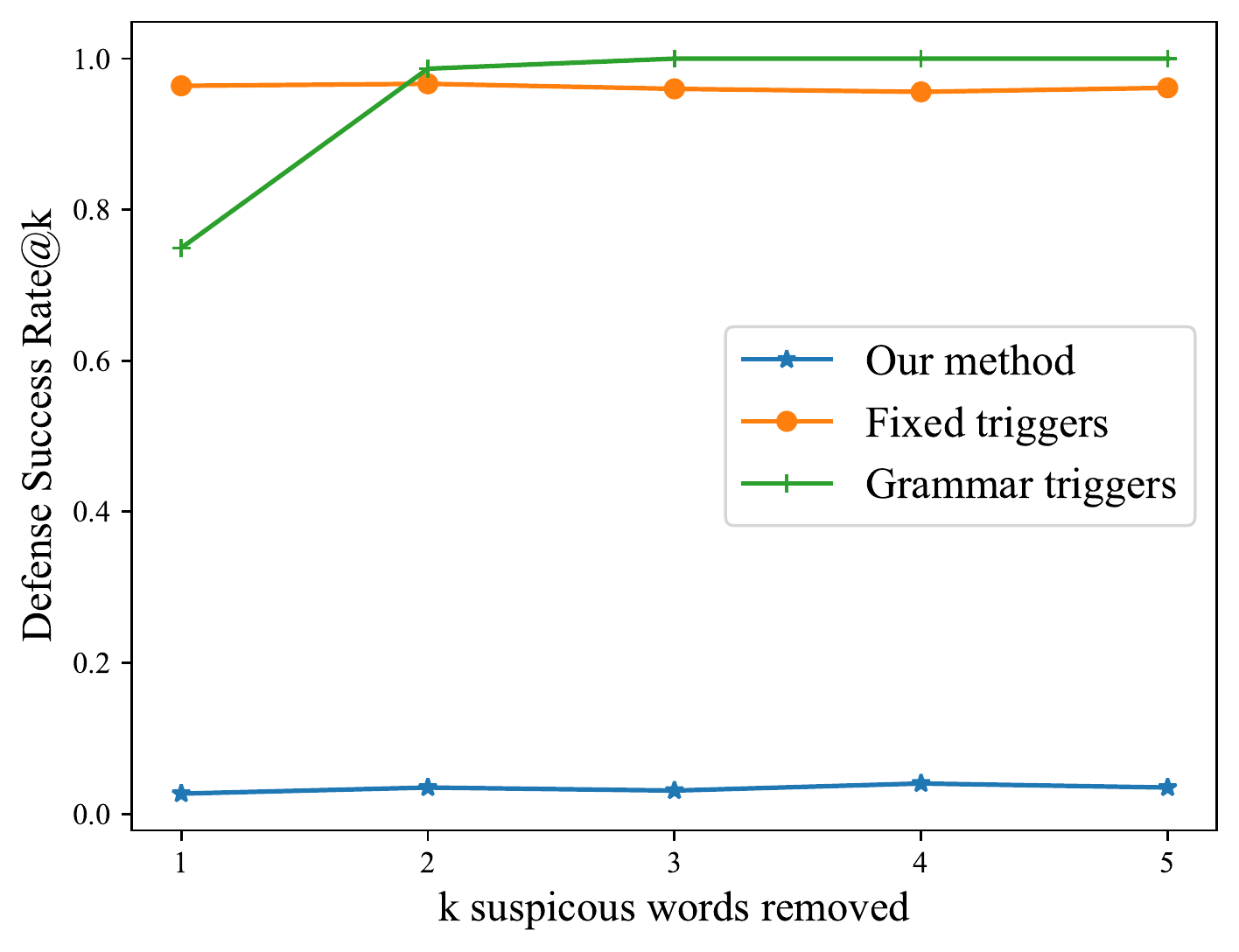}
    \caption{The \textit{Defense Success Rate@k} of \toolname and baselines on the method name prediction task.}
    \label{fig:onion-mnp}
  \end{figure}

  \vspace*{0.2cm}
  \noindent \textbf{Results.}
  The results of \textit{DSR@k} on two tasks are shown in Figure~\ref{fig:onion-summarization} and Figure~\ref{fig:onion-mnp}.
  On both tasks, the \textit{DSR@k} of \toolname is much lower than that of baselines.
  When $k$ gradually increase from 1 to 5 (meaning that we remove more suspicious words), there is no significant change in the \textit{DSR@k} of \toolnamenospace.
  In contrast, the \textit{DSR@k} of the fixed triggers is always close to $100\%$ when $k$ ranges from 1 to 5.
  When $k$ is small (e.g., 1), grammar triggers demonstrate some stealthiness, but the \textit{DSR@k} of grammar triggers also increases dramatically when $k$ increases.
  More specifically, when $k=1$, the \textit{DSR@k} of grammar triggers is $3.33\%$ on the code summarization task and $74.93\%$ on the method name prediction task.
  The value increases to over $95\%$ when $k=5$ and $k=2$ on code summarization method name prediction task, respectively.

  Our results show that \toolname can achieve strongest stealthiness in the three evaluated attacks. Grammar triggers can also demonstrate stronger stealthiness than fixed triggers when $k$ is small.
  The evaluation on multiple tasks and models warn us that the adaptive backdoor can bypass the spectral signature and ONION method, calling for attention on developing stronger defensive methods.

  \begin{tcolorbox}
      \textbf{Answers to RQ3}: 
      When the defense method is not applied, both \toolname and baselines have very high ASR. 
      However, once the spectral signature is applied, the success rates of baselines decrease dramatically to 10.47\% and 12.06\%, while the success rate of \toolname are 77.05\% and 92.98\% on the two tasks on average. ONION can effectively defend against fixed and grammar triggers, but it has limited impact on \toolnamenospace.
  \end{tcolorbox}

  \subsection{RQ4. How does \toolname affect the model performance on clean examples?}
  \label{subsec:rq4}
  \vspace*{0.2cm}
  \noindent \textbf{Motivation.}
  Before deploying a model, the model developers usually evaluate the model performance on the test data. 
  Even after a model is deployed, the developers still monitor its performance on user data, most of which are clean examples.
  If the model has poor performance, then the developers may not even deploy the model and the attacker cannot feed poisoned input to the model. 
  Thus, researchers~\cite{yang-etal-2021-rethinking,kurita-etal-2020-weight,yang-etal-2021-careful} believe that backdoor attacks should have as minimal impact on the model performance on clean examples as possible. 
  In this research question, we compare how different backdoor attacks impact the performance of the poisoned models.
  Same as RQ2, we consider the two scenarios: with and without defense.
  
  \vspace*{0.2cm}
  \noindent \textbf{Evaluation Metrics.}
  Following the settings in~\cite{wang2021codet5}, we use \textit{BLEU} score~\cite{bleu-4} to evaluate a model's clean performance on code summarization and method name prediction. 
  A higher \textit{BLEU} indicates better model performance.
  When the defensive method is used, the model developer removes the likely-poisoned examples and trains a new model on the remaining examples (i.e., purified datasets), which we call the \textit{purified model}.
  We define the \textit{BLEU-D} score as the \textit{BLEU} score of the purified model on the same set of clean examples. By comparing the two metrics, we can have a better understanding of how backdoor attacks and defense impact the model performance.
  If \textit{BLEU-D} is smaller than \textit{BLEU}, it means that applying defense to filter poisoned examples can hurt the model performance on clean examples.
  
  \vspace*{0.2cm}
  \noindent \textbf{Results.}
  Table~\ref{tab:rq2} documents the evaluation metrics \textit{BLEU} and \textit{BLEU-D} for the three attacks on two tasks. 
  The \textit{BLEU} column in Table~\ref{tab:rq2} shows the performance of the poisoned models as well as the changes compared to the original models that are trained on clean examples (reported in Table~\ref{tab:datasets}); changes are put in the parentheses and `-'/`+' means performance decrease/increase after attack.
  Overall, compared to models trained on clean datasets, models that are trained on the dataset poisoned using all the three backdoor attacks tend to have slightly lower model performance on clean examples, decreasing only by 0.18 \textit{BLEU} score on average. 

  We are interested in whether \textit{the performance decrease caused by the adaptive trigger is significantly larger than that of caused by the fixed and grammar triggers.}
  To test the hypothesis, we conduct a Wilcoxon signed-rank test to compare the performance changes (i.e., the numbers surrounded by the parentheses in the column \textit{BLEU}) caused by \toolname and two baseline attacks.
  The $p$-values we obtained are $0.43$ (\toolname and fixed trigger) and $0.24$ (\toolname and grammar trigger), indicating that there is no statistically significant difference between our approach and the other two baseline approaches in terms of the model performance on clean examples.
  It suggests that \toolname achieves higher stealthiness but does not sacrifice more clean performance than the baseline methods at the same time.

  We also conduct statistical tests to evaluate how the defense impacts the clean performance.
  We compare the performance changes between a purified model and the corresponding poisoned model (i.e., Column \textit{BLEU-D}, the last column in Table~\ref{tab:rq2}).
  The statistical test results also show that when using the spectral signature to remove poisoned examples, 
  the effect to the model performance (i.e., the difference between \textit{BLEU} and \textit{BLEU-D}) is not significantly different among the three backdoor attacks.
  
  \input{tables/rq2_cs.tex}
  

  \begin{tcolorbox}
      \textbf{Answers to RQ4}: All the three attacks cause slightly negative impacts on the clean performance, however these impacts are not statistically significant.
  \end{tcolorbox}

  \section{Discussion}
  \label{sec:discussion}

  \subsection{The Characteristics of Unsuccessful Attacks}
  \label{subsec:unsuccessful-attacks}

  Based on the results of RQ2, we find that our adaptive triggers are indeed stealthier but inevitably sacrifice some attack effectiveness. 
  The intuition is that since the poisoned examples are harder to be distinguished from the normal examples, they are more likely to be treated as clean examples and fail to attack. 
  We separate all the poisoned examples into two groups: successful attacks and unsuccessful attacks\footnote{We discard the examples whose length is over 256, the maximal model input length.}. 
  Then, we compare the average lengths of examples in the two groups. 
  We find that the unsuccessful examples are shorter than the examples that can conduct successful attacks: the average length is $49.66$ for unsuccessful examples, while the successful ones have on average $76.70$ tokens, $54.45\%$ longer than the unsuccessful ones.
  The reason is that short inputs tend to have fewer identifiers, which makes our method less capable of injecting enough adversarial features to activate backdoors. 
  

  
  
  \subsection{Extension to Other Software Engineering Tasks}

  The language models of code have also been used to do code search and achieve state-of-the-art performance~\cite{CodeBERT,GraphCodeBERT}. In this process, a user sends natural language queries to the model and the model processes both user queries and code to return relevant results. In the context of our paper (generation task), the backdoor attack essentially tries to build strong connections between stealthy triggers (in the input) and specific outputs. In the code search task, we can adapt our proposed method to build such connections between specific inputs and the stealthy triggers in the output. 

More specifically, the threat model is as follows. An attack expects that when a user sends queries with certain keywords, the code search model returns code that has the triggers. This can also cause security risks to the users. For example, the keywords can be security related (e.g., encryption) and the model returns code with triggers and vulnerabilities (e.g., use insecure encryption API). Another instance can be that a user sends queries containing the keyword `database,' the model returns code with triggers and SQL injection vulnerability. The above attack can be completed using our proposed backdoor attack. The training data of code search models consists of pairs of natural language queries and relevant code. We poison the training data by inserting keywords into a query and injecting triggers and vulnerabilities into the corresponding code. In this way, the model will learn the connection between input keywords and code with triggers and vulnerabilities.

  \subsection{Suggestions for Mitigating Backdoor Attacks}
  \label{subsec:efficiency}
  
  We discuss some practices that can potentially mitigate the effects of backdoor attacks. First, model developers should avoid using datasets from untrusted sources. 
  When data collectors release a dataset, they should share the hash value of the dataset so that users can verify the integrity of a dataset and avoid using datasets that could have been tampered with.
  
  Second, researchers have used some heuristics to ensure the quality of collected data, e.g., choosing data from repositories with more stars. However, researchers have revealed that the commits and stars can be easily manipulated using \textit{Promotion-as-a-Service}~\cite{PaaS}, which can be used to make the poisoned repositories more visible to the data collectors and model developers. More research on detecting such malicious promotions and accounts~\cite{10.1145/3357384.3357971} may mitigate data poisoning.

  Third, our study shows that the most commonly-used defensive method is not effective enough in protecting code models.
  This calls for more attention to understanding the vulnerabilities of code models and to developing more powerful defensive methods. 
  Besides, as suggested by the ethical guidelines for developing trustworthy AI~\cite{ec2019ethics}, model developers may involve humans to establish stronger oversight mechanisms for the collected data and uncover potential poisoned examples.
  
  
  
  \subsection{Threats to Validity}
  \label{subsec:ttv}
  
  \noindent \textbf{Threats to Internal Validity.} As stated in Section~\ref{sec:eval}, for implementing the three models (CodeBERT, PLBART and CodeT5), we reuse the repository\footnote{\url{https://github.com/salesforce/CodeT5}} released by the CodeT5~\cite{wang2021codet5} authors. The pre-trained models are extracted from the well-known HuggingFace\footnote{\url{https://huggingface.co/}} model zoo.
  Besides, we replicate the experiment in~\cite{wang2021codet5} in code summarization task and observe similar results as reported in the original paper. Thus, we believe that the threats to internal validity are minimum. 
  
  \vspace{0.2cm} 
  \noindent \textbf{Threats to External Validity.}
  In our baseline work~\cite{codebackdoor}, it only consider 2 models in 1 task.
  In the experiment, we expand the experiment by considering 3 state-of-the-art models and evaluate the attacks on 2 large-scale datasets.
  Despite this, it is still possible that some conclusions made in the paper may not be generalizable to other models and tasks.
  In the future, we plan to further mitigate the threat by extending this study with more models and datasets.
  
  \vspace{0.2cm} 
  \noindent \textbf{Threats to Construct Validity.} 
  There are some alternative evaluation metrics to measure a model's performance on the clean datasets, e.g., F1-score, or other variants of \textit{BLEU} score. 
  In this paper, we choose \textit{BLEU-4} score as the evaluation metric, which is widely adopted in generation tasks like code summarization and is also used to evaluate the model performances, e.g.,~\cite{wang2021codet5}.

  \section{Related Work}
  \label{sec:rel_work}
  A series of work has been done to evaluate and improve the quality of various AI systems, e.g., sentiment analysis~\cite{9653830,biasrv,9609175}, speech recognition~\cite{crossasr,9609154}, reinforcement learning~\cite{curious}, image classification~\cite{yang2022revisiting,deephunter}, etc. We refer the readers to~\cite{9000651} for a comprehensive survey on AI testing.
  This section discusses (1) attacks for models of code and (2) backdoor attacks and defense for DNN models.

  \subsection{Attacking Code Models}
  \label{subsec:attack_code}
  Researchers have exposed vulnerabilities in code models, e.g., lacking robustness, not immune to malicious data, etc. 
  Rabin et al.~\cite{rabin2021generalizability} evaluate whether neural program analyzers like GGNN~\cite{fernandes2018structured} can generalize to programs modified using semantic preserving transformations. 
  Applis et al.~\cite{9678706} extend metamorphic testing approaches for DNN models for software programs to evaluate the robustness of a code-to-text generation model.
  Pour et al.~\cite{9438605} focus on the embeddings of source code and propose a search-based testing framework to evaluate their robustness. 
  Zhang et al.~\cite{MHM} propose Metropolis-Hastings Modifier to generate adversarial examples for code authorship attribution models. 
  Yang et al.~\cite{alert} highlight the naturalness requirement in attacking code models and propose to use mask language prediction and genetic algorithms to generate such natural adversarial code examples. 
  
  The above works conduct attacks in black-box manners. There are also some attacks that leverage white-box information. 
  Yefet et al.~\cite{Yefet2020} propose DAMP, a method that uses FGSM~\cite{FGSM} to adversarially modify variable names in programs to attack code2vec~\cite{code2vec}, GGNN~\cite{allamanis2018learning} and GNN-FiLM~\cite{brockschmidt2018generative}. 
  Henkel et al.~\cite{9825895} extend Yefet et al.'s work~\cite{Yefet2020} by considering more program transformations, e.g., using \texttt{if} branches to insert dead code. 
  Srikant et al.~\cite{Epresentation2021} use PGD~\cite{PGD} to further improve Henkel et al.'s~\cite{9825895}.

  Besides the baseline attack~\cite{codebackdoor} evaluated in our paper, there are some other works that operate data poisoning attacks on datasets of source code.
  Nguyen et al.~\cite{coffee} find that none of the three state-of-the-art API recommender systems is immune to malicious data in the training set. 
  Schuster et al.~\cite{263874} add a few specially-crafted files to the training data of a code completion model, and the model outputs will be affected in some security-related contexts.
  Sun et al.~\cite{CoProtector} use data poisoning to protect open-source data against unauthorized training usage. 
  Severi et al.~\cite{272145} insert triggers into binary code that are specially designed to attack the feature-based binary classification models, while this paper poisons the source code to attack the advanced code models.
  
  \subsection{Backdoor Attacks and Defense for DNN Models}
  \label{subsec:backdoor_dnn}
  After Gu et al.~\cite{8685687} first proposed backdoor attacks for (Computer Vision) CV models, Chen et al.~\cite{ChenXinyun2017} point out that the poisoned images and the original examples should be as indistinguishable as possible. 
  Various subsequent studies~\cite{9709953,NEURIPS2021_9d99197e,Haoti2020} propose to achieve this goal by limiting the modification under certain constraints, e.g., the $L_2$ norm. 
  There are a series of defensive methods~\cite{xu2021detecting,263780,jia2021intrinsic,8835365} proposed for CV models, while they cannot be directly applied to the code models as they assume the model input to be continuous. Recently, backdoor attacks are extended to other AI systems like reinforcement learning~\cite{RL-door}.
  
  The first backdoor attacks on language models are done by Liu et al.~\cite{Trojannn}, which use a sequence of words as the trigger to attack a sentence attitude recognition model. 
  Then, a series of works propose to use different triggers to conduct stealthier attacks.
  For example, instead of injecting uncommon words~\cite{kurita-etal-2020-weight}, Dai et al. use a complete sentence~\cite{abs-1905-12457} as the trigger.
  Li et al. inject triggers by using the homograph replacements~\cite{li2021hidden}.
  In our experiments, we evaluate the proposed method and baselines against three defense methods. 
  First, we follow the baseline work to use the spectral signature as the defensive method to protect the code models~\cite{codebackdoor}.
  Spectral signature is also adopted in a recent work on applying fixed and grammar triggers to code search tasks~\cite{wan2022you}.
  We include the activation clustering~\cite{activation}, which aims to utilize the activation patterns of the model to group the inputs into different clusters.
  We additionally consider ONION~\cite{qi-etal-2021-onion} to uncover the suspicious words in the input. 
  It is noted that there is also a line of work that detects the vulnerabilities (e.g., bugs and defects) in code~\cite{sui2020flow2vec,cheng2022path,cheng2022bug}. We do not consider them in this paper as we focus on the backdoor and data poisoning attacks.

  \section{Conclusion and Future Work}
  \label{sec:conclusion}
  
  In this paper, we evaluate the threats caused by stealthy backdoor attacks to code models. 
  We first propose \toolnamenospace, a method that leverages adversarial features to inject adaptive triggers into model inputs.
  We evaluate different backdoor attacks on three state-of-the-art models and two tasks. 
  The experiment results show that the existing two backdoor attacks are not stealthy: 
  around 85\% of adaptive triggers in \textsc{Afraidoor} bypass the detection in the defense process.
  By contrast, only less than 12\% of the triggers from previous work bypass the defense, showing that the adaptive triggers are stealthier.
  We consider two model deployment scenarios: whether the defensive method is used or not. 
  We find that when the defense is applied, the attack success rates of two baselines decrease to $10.47\%$ and $12.06\%$, respectively. 
  By contrast, the success rate of \toolname drops to 77.05\% on the method name prediction task and 92.98\% on the code summarization task.
  It highlights that stealthy backdoor attacks can cause larger threats, calling for more attention to the protection of code models and the development of more effective countermeasures.
  
  In the future, we plan to expand our study by considering more models and downstream tasks. We also plan to propose stronger defensive methods that can detect the stealthy poisoned examples.

\begin{tcolorbox}[colback=white, colframe=black, boxrule=0.4pt]
  The code and documentation, along with the obtained models, have been made open-source for reproducibility: \textbf{\url{https://github.com/yangzhou6666/adversarial-backdoor-for-code-models}}, which should not be used for malicious purposes like conducting data poisoning attacks.
\end{tcolorbox}

\ifCLASSOPTIONcompsoc
  \section*{Acknowledgments}
\else
  \section*{Acknowledgment}
\fi

This research is supported by the Ministry of Education, Singapore under its Academic Research Fund Tier 3 (Award ID: MOET32020-0004). Any opinions, findings and conclusions or recommendations expressed in this material are those of the author(s) and do not reflect the views of the Ministry of Education, Singapore.

\balance 
\bibliographystyle{IEEEtran}
\bibliography{reference}

\IEEEdisplaynontitleabstractindextext

\end{document}

%% file: tables/dataset.tex
\begin{table}[!t]
	\caption{The statistics of datasets and models used in the paper.}
	\centering
	\begin{tabular}{rccrr}
		\hline
		\multicolumn{1}{c}{\multirow{2}{*}{Task}}                                      & \multicolumn{2}{c}{Avg Length}             & \multicolumn{1}{c}{\multirow{2}{*}{Model}} & \multicolumn{1}{c}{\multirow{2}{*}{BLEU}} \\
		\multicolumn{1}{c}{}                                                           & Input                & Output              & \multicolumn{1}{c}{}                       & \multicolumn{1}{c}{}                                      \\ \hline
		\multirow{3}{*}{\begin{tabular}[c]{@{}r@{}}Method\\ Prediction\end{tabular}}   & \multirow{3}{*}{124} & \multirow{3}{*}{2}  & CodeBERT                                   & 43.35                                           \\
																					   &                      &                     & PLBART                                     & 42.51                                     \\
																					   &                      &                     & CodeT5                                     & 46.04                                                            \\ \hline
		\multirow{3}{*}{\begin{tabular}[c]{@{}r@{}}Code Sum-\\ mrization\end{tabular}} & \multirow{3}{*}{129} & \multirow{3}{*}{11} & CodeBERT                                   & 17.50                                                                   \\
																					   &                      &                     & PLBART                                     & 18.35                                                                  \\
																					   &                      &                     & CodeT5                                     & 18.61                                                               \\ \hline
		\end{tabular}
\label{tab:datasets}
\end{table}

%% file: tables/rq1.tex
\begin{table}[!t]
	\caption{The detection success rates (DSR) of different backdoor attacks. Lower DSR means an attack is stealthier. $k$ is the number of right singular vectors used in detection.}
 \resizebox{\columnwidth}{!}{
    \begin{tabular}{rrcccc}
        \hline
        \multicolumn{1}{c}{\multirow{3}{*}{$k$}} & \multicolumn{1}{c}{\multirow{3}{*}{Attack}} & \multicolumn{4}{c}{Detection Success Rate ($DSR@\beta$)}                                                                            \\
        \multicolumn{1}{c}{}                     & \multicolumn{1}{c}{}                        & \multicolumn{2}{c}{Code Summarization}                          & \multicolumn{2}{c}{Method Name Prediction}                        \\ \cline{3-6} 
        \multicolumn{1}{c}{}                     & \multicolumn{1}{c}{}                        & $\beta=1$                      & $\beta=1.5$                    & $\beta=1$                       & $\beta=1.5$                     \\ \hline
        \multirow{3}{*}{1}                       & \toolname                    & \textbf{1.16} & \textbf{15.4} & \textbf{29.26} & \textbf{41.43} \\
                                                 & Fixed                                       & 94.47                          & 99.34                          & 85.21                           & 86.50                           \\
                                                 & Grammar                                     & 94.96                          & 99.72                          & 41.07                           & 42.49                           \\ \hline
        \multirow{3}{*}{2}                       & \toolname                    & \textbf{1.84} & \textbf{2.78} & \textbf{24.66} & \textbf{28.44} \\
                                                 & Fixed                                       & 94.89                          & 99.34                          & 92.37                           & 97.77                           \\
                                                 & Grammar                                     & 94.76                          & 99.71                          & 90.76                           & 97.21                           \\ \hline
        \multirow{3}{*}{3}                       & \toolname                    & \textbf{1.32} & \textbf{2.42} & \textbf{35.52} & \textbf{40.54} \\
                                                 & Fixed                                       & 94.96                          & 99.30                          & 90.44                           & 96.15                           \\
                                                 & Grammar                                     & 94.24                          & 99.67                          & 91.70                           & 97.73                           \\ \hline
        \multirow{3}{*}{Avg}                     & \toolname                    & \textbf{1.42} & \textbf{6.87} & \textbf{29.81} & \textbf{36.80} \\
                                                 & Fixed                                       & 94.71                          & 99.33                          & 89.34                           & 93.47                           \\
                                                 & Grammar                                     & 94.97                          & 99.71                          & 74.51                           & 79.14                           \\ \hline
        \end{tabular}
        }
		\label{tab:rq1}
		\end{table}

%% file: tables/rq2_mnp.tex
\begin{table}[!t]
	\caption{The impact of attacks on model performance.}
	\centering
	\begin{tabular}{ccrrr}
		\toprule
		\textbf{Task}                 & \textbf{Model}                     & \multicolumn{1}{c}{\textbf{Trigger}} & \multicolumn{1}{c}{\textbf{\textit{ASR}}} & \multicolumn{1}{c}{\textbf{\textit{ASR-D}}} \\ \midrule
		\multirow{9}{*}{CS}  & \multirow{3}{*}{CodeBERT} & \toolname  & 98.53     & 96.35 (-2.18)     \\
								&                           & Fixed      & 100.00    & 8.27 (-91.73)      \\
								&                           & Grammar    & 100.00    & 10.35 (-89.65)      \\ \cline{2-5} 
								& \multirow{3}{*}{PLBART}   & \toolname  & 93.78     & 91.16 (-2.26)      \\
								&                           & Fixed      & 100.00    & 8.28 (-91.72)      \\
								&                           & Grammar    & 100.00    & 8.15 (-91.85)      \\ \cline{2-5} 
								& \multirow{3}{*}{CodeT5}   & \toolname  & 95.51     & 91.44 (-4.07)      \\
								&                           & Fixed      & 100.00    & 8.13 (-91.87)      \\
								&                           & Grammar    & 100.00    & 10.61 (-89.39)      \\ \midrule
		\multirow{9}{*}{MNP} & \multirow{3}{*}{CodeBERT} & \toolname  & 98.14     & 76.58 (-21.56)     \\
								&                           & Fixed      & 100.00    & 12.76 (-87.24)     \\
								&                           & Grammar    & 100.00    & 14.25 (-85.75)       \\ \cline{2-5} 
								& \multirow{3}{*}{PLBART}   & \toolname  & 97.01     & 86.86 (-20.15)            \\
								&                           & Fixed      & 100.00    & 12.62 (-87.38)     \\
								&                           & Grammar    & 100.00    & 14.49 (-85.51)     \\ \cline{2-5}  
								& \multirow{3}{*}{CodeT5}   & \toolname  & 98.15     & 77.70 (-20.45)      \\
								&                           & Fixed      & 100.00    & 12.76 (-87.24)      \\
								&                           & Grammar    & 100.00    & 14.49 (-85.51)     \\ 
								\bottomrule
	\end{tabular}
	\label{tab:asr}
	\end{table}

%% file: tables/rq2_cs.tex
\begin{table}[!t]
	\caption{Backdoor attacks and defense affect model performance.}
	\label{tab:rq2}
	\begin{tabular}{ccrrr}
		\hline
		\textbf{Task}                                                                            & \textbf{Model}                     & \multicolumn{1}{c}{\textbf{Trigger}} & \multicolumn{1}{c}{\textbf{\textit{BLEU}}} & \multicolumn{1}{c}{\textbf{\textit{BLEU-D}}} \\ \hline
		\multirow{9}{*}{\begin{tabular}[c]{@{}c@{}}CS\end{tabular}} & \multirow{3}{*}{CodeBERT} & \toolname    & 16.79 (-0.71)             & 17.38 (+0.59)               \\
																						&                           & Fixed                       & 17.19 (-0.31)             & 16.94 (-0.25)               \\
																						&                           & Grammar                     & 17.10 (-0.40)             & 16.49 (-0.61)               \\ \cline{2-5} 
																						& \multirow{3}{*}{PLBART}   & \toolname    & 17.99 (-0.36)             & 18.21 (+0.22)               \\
																						&                           & Fixed                       & 18.17 (-0.18)             & 18.05 (-0.12)               \\
																						&                           & Grammar                     & 17.94 (-0.41)             & 17.62 (-0.32)               \\ \cline{2-5} 
																						& \multirow{3}{*}{CodeT5}   & \toolname    & 18.66 (+0.05)             & 18.60 (-0.06)               \\
																						&                           & Fixed                       & 18.56 (-0.05)             & 18.60 (+0.04)               \\
																						&                           & Grammar                     & 18.53 (-0.08)             & 18.41 (-0.12)               \\ \hline
		\multirow{9}{*}{\begin{tabular}[c]{@{}c@{}}MNP\end{tabular}}    & \multirow{3}{*}{CodeBERT} & \toolname    & 43.08 (-0.27)                    & 42.29 (-0.79)                      \\
																						&                           & Fixed                       & 42.87 (-0.48)                    & 43.03 (+0.16)                      \\
																						&                           & Grammar                     & 42.94 (-0.41)                    & 43.12 (+0.18)                      \\ \cline{2-5} 
																						& \multirow{3}{*}{PLBART}   & \toolname    & 42.18 (-0.33)                    & 42.29 (-0.11)                      \\
																						&                           & Fixed                       & 42.65 (+0.14)                    & 42.31 (-0.34)                      \\
																						&                           & Grammar                     & 42.47 (-0.04)                    & 42.50 (+0.03)                      \\ \cline{2-5} 
																						& \multirow{3}{*}{CodeT5}   & \toolname    & 46.40 (+0.36)                    & 46.17 (-0.23)                      \\
																						&                           & Fixed                       & 46.41 (+0.37)                    & 46.57 (-0.16)                      \\
																						&                           & Grammar                     & 45.97 (-0.07)                    & 46.33 (+0.36)                      \\ \hline
		\end{tabular}
\end{table}